\documentclass[onecolumn,showpacs,nobibnotes,nofootinbib,12pt,longbibliography]{revtex4-1}
\usepackage{amsmath,amssymb}
\usepackage{wasysym}
\usepackage{graphicx}
\usepackage{multirow}
\usepackage{color}
\usepackage[colorlinks,linkcolor=blue,citecolor=blue]{hyperref}

\newcommand{\be}{\begin{equation}}
\newcommand{\ee}{\end{equation}}
\newcommand{\bea}{\begin{eqnarray}}
\newcommand{\ea}{\end{eqnarray}}

\newcommand{\nn}{\nonumber\\}

\newcommand{\f}{\frac} 
\def\d{\partial}

\def\be{\begin{eqnarray}}
\def\ee{\end{eqnarray}}
\def\st{\begin{equation}}
\def\stp{\end{equation}}

\def\p{{\bf p}}
\def\x{{\bf x}}
\def\v{{\bf v}}

\def\M{\mathcal{M}}
\def\C{\mathcal{C}}
\def\I{\mathcal{I}}
\def\L{\mathcal{L}}
\def\J{\mathcal{J}}
\def\F{\mathcal{F}}
\def\S{\mathcal{S}}

\def\Eq#1{eq.~(\ref{#1})}
\def\Eqs#1{eqs.~(\ref{#1})}
\def\app#1{Appendix~\ref{#1}}
\def\Fig#1{Fig.~\ref{#1}}
\def\Sect#1{Section~\ref{#1}}
\def\Tab#1{Table~\ref{#1}}

\def\eg{{\it e.g.}}

\begin{document}

\title{Quark production, Bose-Einstein condensates and thermalization of the quark-gluon plasma}
\author{Jean-Paul Blaizot}
\author{Bin Wu}
\author{Li Yan}
\affiliation{Institut de Physique Th\'{e}orique, CEA Saclay, 91191, Gif-sur-Yvette Cedex, France}
\begin{abstract}
In this paper, we study the thermalization of gluons and $N_f$ flavors of massless quarks and antiquarks in a spatially homogeneous 
system. First, two coupled transport equations for gluons and quarks (and antiquarks) are derived within the diffusion approximation of 
the Boltzmann equation, with only $2\leftrightarrow 2$ processes included in the collision term. Then, these transport equations are solved 
numerically in order to study the thermalization of the quark-gluon plasma. At initial time, we assume that only gluons are present 
and we choose the gluon distribution of a  form inspired by the color glass picture, namely $f = f_0~\theta\left(1-\f{p}{Q_s} \right) $ with $Q_s$ the saturation momentum and 
$f_0$ a constant. The subsequent evolution of the system
may, or may not, lead to the formation of a (transient) Bose condensate (BEC) of gluons, depending on the value of $f_0$. In fact, we observe, depending
on the value of $f_0$, three different patterns:
(a) thermalization 
without BEC for $f_0\leq f_{0t}$, (b) thermalization with transient BEC for $f_{0t} < f_0 \leq f_{0c}$, 
and (c) thermalization with BEC for $f_{0c} < f_0 $. The values of $f_{0t} $ and $f_{0c} $ depend on $N_f$. When $f_0\gtrsim1 > f_{0c}$, the onset  of BEC occurs at a finite time $t_c\sim\f{1}{(\alpha_s f_0)^2}\f{1}{Q_s}$. We also find that quark production slows down the thermalization process: the equilibration time 
for $N_f = 3$ is typically about 5 to 6 times longer than that for $N_f = 0$ at the same $Q_s$ and $f_0$.
\end{abstract}
\maketitle

\section{Introduction}\label{sec:intro}
Understanding how  a dense system of gluons evolves into a thermalized  quark-gluon plasma (QGP)  is an important, and theoretically challenging, problem. After two colliding nuclei pass through each other in a relativistic heavy ion collision (HIC), a dense system of gluons is believed to be produced in a time scale of order $t\sim 1/{Q_s}$, with $Q_s$ the saturation momentum characterizing the initial nuclear wave functions \cite{Mueller:1999:initial}. In this early stage,  $f_0$, the occupation number of the produced gluons with $p\lesssim Q_s$, may be as large as $1/{\alpha_s}$, where $\alpha_s$ is the strong coupling constant. Under such conditions, it has been argued that  a Bose-Einstein condensate (BEC) may develop during the approach to equilibrium, provided inelastic, number changing, processes do not play a too important role \cite{Blaizot:2011xf,Blaizot:2013:BEC}. The effect of such inelastic processes remains a somewhat controversial issue. Of course, number changing  processes exclude the existence of a  BEC in the equilibrium state. The real issue is therefore whether a transient BEC can emerge as the system evolves towards thermalization. Various arguments against this possibility are presented in Ref.~\cite{Kurkela:2012hp}, while the calculations in  Ref.~\cite{Huang:2013:2to3}  suggest that inelastic processes could   amplify the growth of soft gluon modes, thereby accelerating the formation of a BEC \cite{Huang:2013:2to3}. We shall not attempt to resolve this issue here, but consider rather   the effect of another type of inelastic processes leading to the variation in the gluon number, namely processes that involve the creation of quark-antiquark pairs.

The partons that are produced in the early stage of HIC are mostly gluons: the number of quarks and antiquarks is initially negligible compared to the large number of gluons. However, in a thermalized quark-gluon plasma, the energy density is given by 
\be
\epsilon = 3 P = \left[16 + \f{21}{2} N_f \right] \f{\pi^2}{30}T^4,
\ee 
where we have assumed non-interacting quarks and gluons,  $N_f$ flavors of massless quarks (and antiquarks), and  $T$ is the temperature. At the energies of RHIC and LHC, one may take $N_f = 3$. In this case quarks and antiquarks carry $66\%$ of the total energy density. Therefore, the study of quark production in a dense system of gluons is obviously of great importance to fully understand  the thermalization of the quark-gluon plasma.

In this paper, we  obtain two coupled kinetic equations for both gluons and quarks (and antiquarks),  using   the Boltzmann equation in the diffusion approximation \cite{PhysicalKinetics}. The collision term contains all the $2\leftrightarrow 2$ scatterings between quarks and gluons, but only those $2\leftrightarrow 2$ scatterings, with the exclusion of, for instance, inelastic $2\leftrightarrow 3$ processes. We assume the dominance of small angle scatterings which justifies the diffusion approximation.  The baryon number density is assumed to be zero. As a result quarks and antiquarks are described  by the same transport equation, which is coupled to that for gluons. These transport equations are solved numerically to study the thermalization of the quark-gluon plasma. 

The present study complements that carried out in Ref.~\cite{Blaizot:2013:BEC} where quark production was ignored. As in \cite{Blaizot:2013:BEC}, the discussion relies on the Boltzmann equation in the small angle approximation\cite{Mueller:1999:Boltzmann, Venugopalan:2000:Thermalization, Blaizot:2013:BEC}, and both quarks and gluons are taken to be massless. As in \cite{Blaizot:2013:BEC}, we restrict ourselves to the study of a spatially homogeneous non-expanding system. In contrast to Ref.~\cite{Blaizot:2013:BEC}, we are able to follow, albeit approximately, the evolution of the system across  the onset of BEC all the way to thermalization. This is achieved by imposing a specific boundary condition on the solution of the coupled equations at zero momentum.  It is shown in Ref. \cite{Blaizot:2013:BEC} that the formation of BEC starts in an over-populated system at a finite time $t_c$ when the gluon distribution $f$ becomes singular at $p=0$. In this paper, we show that, for $t>t_c$, no solution of the transport equations exists if the total number of partons with $p>0$ is required to be conserved. However, we find solutions by properly imposing a boundary condition that corresponds to a non-vanishing gluon flux at $p=0$. Those solutions are used to describe the evolution of the system beyond $t_c$ all the way to thermal equilibrium, with the number density of condensed particles being deduced from the gluon flux at $p=0$. Note that the procedure just outlined represents presumably a crude approximation to the actual dynamics of particles in the presence of a condensate, but it has the virtue of allowing us to follow continuously the system all the way to its actual thermal equilibrium state.

Quark production decreases the total number of gluons in  the system and could potentially hinder the formation of a BEC. However the $2\leftrightarrow 2$ processes included in the Boltzmann equation conserve the total number of partons. As a result of this conservation law, a chemical potential develops dynamically as the system evolves \footnote{In fact, because the thermalization of quarks proceeds at a slower pace than that of the (soft) gluons, two different chemical potentials develop dynamically, one for the quarks and one for the gluons. These chemical potentials converge to a common value only close to thermalization}. The equilibrium state is achieved for a negative value of this chemical potential, provided the initial number of gluons is not too large. We qualify this situation as under-population. If, on the contrary, the initial population of gluons is large enough, no equilibrium exists without a BEC: this is the situation of over-population, which was found to occur in the absence of quark production, and was thoroughly studied in \cite{Blaizot:2013:BEC}. Thus the present study shows that quark production delays the onset of BEC but does not prevent the occurrence of the phenomenon. In fact, because the growth of the population of soft gluon modes is a fast phenomenon, and quark production is relatively slow, one even encounters situations where a transient BEC appears in the course of the evolution to equilibrium, before being eventually suppressed when quark production takes over and eliminates the excess gluons prior to  thermalization.

The paper is organized as follows. The transport equations for quarks and gluons are derived in Sec. \ref{sec:equations}. In Sec. \ref{sec:equilibrium}, the parameters that characterize the thermodynamic equilibrium are determined from the initial conditions, assuming that the total parton number is fixed. Our main results obtained by solving the transport equations for various type of initial conditions are presented in Sec. \ref{sec:results}. We conclude in Sec. \ref{sec:discussions}. Appendix \ref{app:M2} gives some details about the derivation of the transport equations. In Appendix \ref{app:series}, we present series solutions of the transport equations that are valid at small $p$. These are used in particular to set appropriate boundary conditions at $p=0$ in the various regimes encountered. 
\section{Transport equations for a quark-gluon system}
\label{sec:equations} 
The analysis, in the framework of kinetic theory, of the evolution of a quark-gluon system towards equilibrium relies on the 
possibility to describe quark and gluon degrees of freedom in terms of phase space distributions. Color and spin degrees of freedom 
do not play essential roles in the present discussion and they
will be averaged out. We shall denote the color and spin averaged distribution function of gluons with 
$f(t, \x,\p)$ and that of quarks with $F(t, \x,\p)$ throughout this paper, except in  very few cases, such as in \Eqs{boltz0} or (\ref{collision}) below, when a different notation is found more convenient.

In this section we obtain two coupled transport equations that govern the evolution of $f(t, \x,\p)$ and $F(t, \x, \p)$. In a thermal bath of quarks and gluons, the number density of quarks whose masses are much heavier than the 
temperature $T$ is negligibly small compared to that of light quarks and gluons. We thus only consider 
the $N_f$ flavors of quarks, and their antiparticles, whose masses are smaller than $T$, and take them to be massless for simplicity. Furthermore, 
we assume that the baryon number density is zero everywhere in the system, and no external forces are exerted on the partons. In this case,  quarks and antiquarks have the same 
distribution due to the $SU(N_f)$ flavor symmetry and  the charge conjugation invariance of QCD. Therefore, one only needs two 
coupled equations for the quark distribution $F$ and the gluon distribution $f$ to describe the evolution of the system.

Although the number of colors and of flavors are both commonly taken to be $N_c=N_f=3$ in realistic phenomenological studies of heavy-ion collisions, we will keep here $N_c$ and 
$N_f$ as free parameters. 

\subsection{The Boltzmann equation in the diffusion approximation}
\label{sec:boltz}
The Boltzmann equation
\be
\label{boltz0}
D_t f_{\bf p}^a \equiv \left(\frac{\d}{\d t}+\bf v \cdot \nabla_{\bf x} \right) f_{\bf p}^a = \C[f_{\bf p}^a]\,,
\ee
describes the evolution of the phase space distribution function $f_{\bf p}^a$ with the collision term 
$\C[f_{\bf p}^a]$, including all the $2\leftrightarrow 2$ scattering processes in QCD, of the form 
\begin{align}
\label{collision}
\C[f_\p^a]=&\frac{1}{2 E_p \nu_a}\sum\limits_{b,c,d}\f{1}{s_{cd}}\int \frac{d^3\p'}{(2\pi)^32 E_{\p'}}
\frac{d^3{\bf k}}{(2\pi)^3 2E_{\bf k}}\frac{d^3{\bf k'}}{(2\pi)^32E_{\bf k'}}
(2\pi)^4\delta^{(4)}(P+P'-K-K')
|\mathcal{ M}_{cd}^{ab}|^2 \nonumber\\
&\times\left[f_{\bf k}^cf_{\bf k'}^d(1 + \epsilon_a f_{\bf p}^a)(1+ \epsilon_b f_{\bf p'}^b)- f_{\bf p}^a f_{\bf p'}^b(1 + \epsilon_c f_{\bf k}^c)(1+ \epsilon_d f_{\bf k'}^d))\right]\,,
\end{align}
where a short-hand notation $f_\p^a$ is used for the distribution function of different species with the superscript $a$ distinguishing the different particles. Capital letters are used to denote a four-vector, \eg, the 
four-momentum $P$. Correspondingly, the small and bold letter $\p$ is used for the three vector, while small ordinary letter $p$
stands for its module. The symbol $\epsilon_a$ distinguishes fermions and bosons: 
$\epsilon_a = 1$ for bosons and $\epsilon_a = -1$ for fermions.
In \Eq{collision}, the color and spin degrees of freedom of incoming particles $a$ and $b$, 
and the outgoing particles $c$ and $d$, have been summed over in the squared scattering matrix element $|\mathcal{ M}_{cd}^{ab}|^2$. 
The factor $\nu_a$ stands for the number of spin $\times$ color degrees of freedom of particle
$a$ (which is $2(N_c^2-1)$ for a gluon and $2 N_c$ for a quark or an antiquark), and reflects the corresponding averaging of the initial state particle $a$. 
The factor $s_{cd}$ is a symmetry factor: $s_{cd}$ = $2$ if $c$ and $d$ are identical particles and $s_{cd} = 1$ otherwise.

In a pure gluon system, the differential cross-section $gg\leftrightarrow gg$ diverges if the momentum transfer $\bf q$ is much smaller 
than the momenta of the two scattering gluons. Thus, low momentum transfer or small angle scatterings dominate, which allows us to treat the Boltzmann equation in a  diffusion approximation. The kinetic equation then  reduces to a Fokker-Planck equation \cite{Mueller:1999:Boltzmann, Blaizot:2013:BEC} 
\be
D_t f = -\nabla_{\p}\cdot\J,
\ee
where $\J$  is  an effective current that summarizes the effect of the (small angle) collisions. This current is proportional to a logarithmically divergent  integral of the form 
\be\label{eq:L}
\L\simeq\int_{q_{min}}^{q_{max}}\frac{d q}{q}\,,
\ee 
where $q_{min}$  is of the order of  the screening  mass, while $q_{max}$ is of the order of the largest typical momentum in the system (e.g. the   temperature  if the system is close to equilibrium \cite{Arnold:2002:Boltzmann}).

In a quark-gluon system, the small angle scatterings between quarks and gluons are also important. These contribute to two  currents: $\J_g$ for gluons and $\J_q$ for quarks.  In addition to the effect of collisions which do not alter the nature of the colliding particles, there are equally important production processes:  $q\bar q \leftrightarrow gg$, $qg\leftrightarrow qg$ and $\bar qg\leftrightarrow \bar qg$.  These production processes of quarks (gluons) in the scattering of gluons (quarks) with other particles  contribute to source terms: $\S_g$ for  the production of gluons, and $\S_q$ for the production of quarks. 

By tracking the dominant contributions from all the $2\leftrightarrow 2$ scattering processes between quarks and gluons, as 
listed in \Tab{qcd2t2} of Appendix \ref{app:M2}, we then obtain two diffusion-like equations
\begin{subequations}
\label{boltz_equ}
\bea
\label{boltz1}
D_t f =&-\nabla_{\p}\cdot \J_g+\S_g,\\
\label{boltz2}
D_t F =&-\nabla_{\p}\cdot \J_q+\S_q,
\ea 
\end{subequations} 
where the currents are given by
\begin{subequations}
\label{qg_currents}
\bea
\label{current_g}
\J_g&=&-4\pi\alpha_s^2 N_c \L\left[\I_a \nabla_{\bf p} f+\I_b \frac{\p}{p} f (1+f )\right]\,,\\
\label{current_q}
\J_q&=&-4\pi\alpha_s^2 C_F \L\left[\I_a\nabla_{\bf p} F+ \I_b \frac{\p}{p} F ( 1-F )\right]\,,
\ea
\end{subequations}
and the sources by
\be
\label{sources}
\S_g=-\frac{N_f}{C_F}\S_q=\frac{4\pi\alpha_s^2  C_F N_f \L\I_c}{ p}\left[ F (1+f ) - f ( 1 - F ) \right]\,,
\ee
with
\begin{subequations}
\label{iabc}
\bea
\label{ia}
\I_a&=&\int \frac{d^3\p}{(2\pi)^3} \left[N_c f (1+f ) +N_f F( 1-F )\right]\,,\\
\label{ib}
\I_b&=&2\int \frac{d^3 \p}{(2\pi)^3} \f{1}{p}\left( N_c f+N_f F\right)\,,\\
\label{ic}
\I_c&=&\int \frac{d^3 \p}{(2\pi)^3} \f{1}{p} ( f+F )\,.
\ea
\end{subequations}
Here, $C_F=(N_c^2-1)/(2N_c)$ is the
square of the Casimir operator of the color $SU(N_c)$ group in the fundamental representation. In the following, we shall often refer to the first and second terms on the right hand side of eqs. (\ref{boltz_equ}) respectively  as the diffusion and source terms, although only the contributions proportional to ${\cal I}_a$ in the currents correspond truly  to diffusion processes. 

\begin{figure}
\begin{center}
\includegraphics[width=0.6\textwidth]{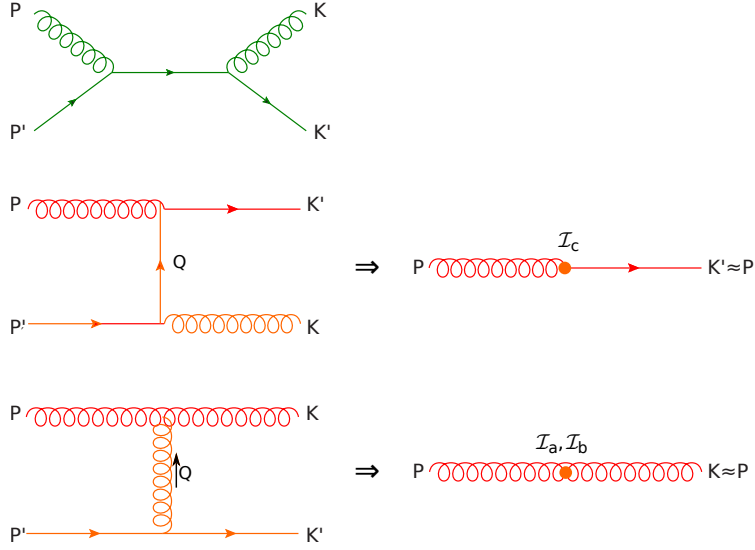}
\end{center}
\caption{(Color online)  Diagrams for $gq\to gq$. In the diffusion approximation, the first diagram can be neglected, the square of the second diagram contributes to the source terms $S_q$ (and $S_g$) and the square of the third diagram contributes to  the current $\J_g$. The diagrams in the right hand side illustrate the fate of the particle that is followed in the Boltzmann equation, chosen here to be a gluon. The diagram in the second line describes  the change of the gluon into a quark (identified as the particle with momentum close to that of the gluon), and its contribution is proportional to  the integral  $\I_c$ which involves integrating over the other two partons in the process. The diagram in the third line describes the diffusion of the gluon in momentum space, which is controlled by the two integrals  $\I_a$ and $\I_b$, which   also  involve integrating out the other two partons, here the two quarks in the lower part of the process displayed on the left.
}\label{fig:gqgq}
\end{figure}

A few comments on these new equations are in order.
First, although we postpone the detailed derivations of \Eqs{boltz_equ} to 
\app{app:M2}, the essential steps and concepts
in these derivations can be revealed by focusing on one of the scattering processes, $gq\to gq$, for example.  
The corresponding diagrams are shown in Fig. \ref{fig:gqgq}, and the square of the associated matrix element is 
\be
\label{gqgq}
| \M_{gq}^{gq}|^2 /g^4= - 8 N_c C_F^2\left(\frac{u}{s}+\frac{s}{u}\right) + 8 N_c^2 C_F \frac{u^2+s^2}{t^2}
\,,
\ee
where the Mandelstam variables are $s=(P+P')^2$, $t=(P - K)^2$ and $u=(P-K')^2$. There are two types of divergent terms in \Eq{gqgq} when $q\rightarrow 0$ (small angle approximation). The $u$ channel term ($\sim 1/u$) comes from the square of 
the second diagram in Fig. \ref{fig:gqgq}, while the $t$ channel term ($\sim 1/t^2$) comes from the square of the third diagram 
in Fig. \ref{fig:gqgq}. Substituting \Eq{gqgq} back into \Eq{collision},
one finds that the two dominant terms are actually of the same order in the logarithmic approximation
\footnote{Here, the difference between the medium-dependent masses of quarks and gluons is neglected, which is valid in the 
leading logarithmic approximation.}.
The $t$ channel scattering results in a part of the currents in \Eqs{qg_currents}, 
while the $u$ channel contributes to the sources,  \Eq{sources}.
Repeating the same analysis for all the other $2\leftrightarrow 2$ scattering processes, we obtain \Eqs{boltz_equ}. 

A second comment is that the reduced collision terms in \Eqs{boltz_equ} preserve 
important physical properties of the original kinetic equation, \Eq{boltz0}. For instance, it can be verified that
the equilibrium Bose-Einstein
distribution for gluons, and Fermi-Dirac distribution for quarks, are still the fixed point solutions to \Eqs{boltz_equ} with a temperature given by $T={\cal I}_a/{\cal I}_b$. Besides, 
the collision terms in the diffusion form conserve energy, and particle number.
We provide an explicit proof for a specified case in the next section. 
\subsection{The transport equations for spatially homogeneous systems}
In the following, we shall study a spatially homogeneous system of quarks and gluons. In this case the spatial dependence 
of the phase space distribution can be ignored and $D_t=\f{\partial}{\partial t}$. In addition, we assume isotropy of the momentum distributions, which are then solely functions of the modulus of the  momentum and of time.
We introduce a new time variable
\be\label{eq:tautot}
\tau = \frac{2 \alpha_s^2 N_c \L}{\pi} t,
\ee
and denote the derivatives with respect to $\tau$ and $p$ by overdots and primes respectively.
Then \Eqs{boltz1} and (\ref{boltz2}) reduce to
\bea
\label{boltz1dg}
&&\dot{f}  = -\frac{1}{p^2} \left( p^2 J_g\right)'+\frac{C_F N_f}{N_c}S_g =  -\frac{1}{4 \pi p^2} \F_g' - \frac{C_F N_f}{N_c}S_q\label{eq:ftoSolve},\\
\label{boltz1dq}
&&\dot{F} =-\f{C_F}{N_c}\frac{1}{p^2} \left( p^2 J_q \right)'+\frac{C_F^2}{N_c} S_q = -\f{C_F}{N_c}\frac{1}{4 \pi p^2} \F_q' + \frac{C_F^2}{N_c} S_q\label{eq:FtoSolve}\,,
\ea
where we have introduced the rescaled currents $J_g$ (for gluons) and $J_q$ (for quarks), together with the corresponding fluxes $\F_g$ and $\F_q$:
\bea
&&\frac{\mathcal{F}_g}{4\pi p^2}\equiv J_g\equiv - I_a f'  -  I_b f (1 + f),\\
&&\frac{\mathcal{F}_q}{4\pi p^2}\equiv J_q\equiv - I_a F' - I_b F (1 - F) ,
\ea
and the rescaled source terms
\bea
S_g = -S_q = \frac{I_c}{p} \left[F (1+f) - f ( 1-F ) \right]\,.
\ea 
In the equations above, the integrals $I_a$, $I_b$ and $I_c$ are defined by
\bea
&&I_a = 2 \pi^2 \I_a = \int_0^\infty dp\, p^2\left[ N_c f (1+f) + N_f F (1-F)\right]\,,\nn
&&I_b = 2 \pi^2 \I_b= 2 \int_0^\infty dp\, p \left(N_c f + N_f F\right), 
\qquad I_c = 2 \pi^2 \I_c= \int_0^\infty dp\, p \left( f + F \right).
\ea

Parton number density $n$, and energy density $\epsilon$, are given in terms of the distribution
functions $f$ and $F$ by
\bea
&&n=4 N_c \int\f{d^3\p}{(2\pi)^3}\left( C_F f + N_f F \right) \equiv n_g + n_q,\label{eq:n}\\
&&\epsilon=4 N_c \int\f{d^3\p}{(2\pi)^3}p\left( C_F f + N_f F \right) \equiv \epsilon_g + \epsilon_q.\label{eq:e}
\ee
In a similar manner, 
the entropy density of gluons $s_g$ and of quarks $s_q$ can be expressed in terms of $f$ and $F$ as
\begin{subequations}
\label{entropy}
\bea
&&s_{g}\equiv - 4 N_c C_F\int \f{d^3\p}{(2\pi)^3}\left[ f \log f - (1+f)\log(1+f) \right],\\
&&s_{q}\equiv - 4 N_c N_f\int \f{d^3\p}{(2\pi)^3}\left[  F \log F  + (1- F)\log(1- F)\right]\,,
\ea
\end{subequations}
with the total entropy density of the quark-gluon system given by
\be
s = s_g+s_q.
\ee
The time evolution of $n$, $\epsilon$ and $s$ can be obtained from eqs. (\ref{eq:ftoSolve}) and (\ref{eq:FtoSolve}). The corresponding equations take the following form
\bea
\dot{n}= &&-\left.\f{1}{2\pi^3} C_F ( N_c\mathcal{F}_g + N_f \mathcal{F}_q  )\right|^{p=\infty}_{p=0}, 
\label{eq:ndt}\\
\dot{\epsilon}
= &&-\left.\f{1}{2\pi^3} C_F  \left[ p ( N_c\mathcal{F}_g + N_f \mathcal{F}_q  )+  I_a 4 \pi p^2(N_c f + N_f F) \right]\right|^{p=\infty}_{p=0}\label{eq:edot},\\
\dot{s}= &&\left.\f{C_F}{2\pi^3}\left[ N_c\left( \mathcal{F}_g \log\f{f}{1+f} - 4 \pi p^2 I_b f \right) + N_f\left( \mathcal{F}_q \log\f{F}{1-F} -  4 \pi p^2 I_b F \right) \right]\right|^{p=\infty}_{p=0}\label{eq:sdot}\nn
&&+\f{2 C_F}{\pi^2}\int_0^\infty dp p {s^+(p)},
\ee
where $s^+(p)$ is the non-negative function, 
\bea
s^+ \equiv&& \f{p }{I_a} \left( \f{N_c J_g^2}{f(1+f)} + \f{N_f J_q^2}{F(1- F)}  \right)\nn
&& + C_F N_f I_c \left[ F (1+f) - f(1-F) \right]\log\f{F(1+f)}{f(1-F)}.
\ee

At this point, it is instructive to discuss the conservation of 
the total number of partons and of the energy, as well as the increase of the entropy. To do so, one needs to know the behavior of $f$ and $F$ 
near $p=0$ (the contributions as $p\rightarrow\infty$ to the time derivatives in eqs. (\ref{eq:ndt}, \ref{eq:edot}, \ref{eq:sdot}) vanish and, therefore, can be dropped).  
As discussed in  Appendix \ref{app:series}, two kinds of solutions near $p=0$ are allowed by the 
transport equations (\ref{eq:ftoSolve}) and (\ref{eq:FtoSolve}). For both types of solutions, the boundary terms on the right 
hand side of eqs. (\ref{eq:edot}) and (\ref{eq:sdot}) always vanish. Therefore, $\dot{\epsilon} = 0$  and $\dot{s}\geq0$. 
However, $n$ is not conserved with both solutions. For solutions in which $f$ and $F$ are analytic near $p=0$, 
$n$ is conserved because $\mathcal{F}_g$ and $\mathcal{F}_q$ vanish at $p=0$. But for solutions of the form
\bea
&&f=\f{c_{-1}}{p}-\f{1}{2}+\cdots,\\
&&F=\f{1}{2}+\cdots,
\ee
there is a non-vanishing gluon flux $\F_g$ at $p=0$
\be\label{eq:fbc}
\mathcal{F}_g|_{p= 0} = 4 \pi c_{-1}(I_a  - I_b c_{-1}) \quad\mbox{and}
\quad \mathcal{F}_q|_{p= 0} = 0\,.
\ee
This entails a time variation of the number density
\be\label{eq:ndot}
\dot{n}= && \f{N_c^2-1}{\pi^2}   I_b  c_{-1} (T^*  - c_{-1})\,,
\ee
where we have set
\be
T^*\equiv \f{I_a}{I_b},
\ee
and the coefficient $c_{-1}$ depends only on $\tau$. The non-vanishing gluon flux at $p=0$ reflects the accumulation of gluons of the 
zero mode, whose number density $N^0$ evolves according to
\be\label{eq:np0t}
\dot{N}^0 = - \dot{n},
\ee
in order to ensure the overall conservation of the parton number. 

In the following, we shall follow Ref. \cite{Blaizot:2013:BEC} and neglect the mild time dependence of $\L$ in eq. (\ref{eq:L}). In this case eqs. (\ref{eq:ftoSolve}) and (\ref{eq:FtoSolve}) are invariant under the following scaling transformation
\be\label{eq:scaling}
Q_s \to c Q_s, \quad \tau\to \f{\tau}{c}, \quad {\bf p} \to c \bf{p}
\ee
with $c>0$. As a result, one can express all momenta in units of $Q_s$ (and the same for the chemical potential $\mu$ and the temperature $T$ of 
\Sect{sec:equilibrium}) and times in units of $1/Q_s$.
\section{Thermodynamics of QGP with a fixed total parton number}
\label{sec:equilibrium}
Since only $2\leftrightarrow 2$ processes are included in the collision term of the transport equations, the total parton 
number is conserved. As a result, in equilibrium, gluons, quarks and antiquarks all have the same chemical potential associated to 
parton number conservation. The thermal equilibrium distributions are the fixed points of eqs. (\ref{eq:ftoSolve}) 
and (\ref{eq:FtoSolve}), and are of the form
\bea\label{eq:feq}
f_{eq} = \frac{1}{e^{(p-\mu)/T}-1}, \qquad F_{eq} = \frac{1}{e^{(p-\mu)/T}+1}.
\ea
In the following $T$ and $\mu$ will always refer to as the thermal equilibrium temperature and chemical potential.

The thermodynamic properties of such a QGP are determined by the total energy density and 
the total parton number density, which are respectively denoted by $\epsilon_{0}$ and  $n_0$.  The under-populated and over-populated 
systems have very different properties\cite{Blaizot:2011xf, Blaizot:2013:BEC}. In an under-populated system, the values of $T$ and $\mu<0$ can be 
obtained by solving the equations
\be\label{eq:muTunder}\epsilon_{eq} = \epsilon_0, \qquad n _{eq} = n_0,\ee
where $n_{eq}$ and $\epsilon_{eq}$ are obtained by plugging $f_{eq}$ and $F_{eq}$ into eqs. (\ref{eq:n}) and (\ref{eq:e}). In an 
over-populated system, $n_0$ is so large such that no real solution to the above equations exists. The thermal distributions are then given 
by $f_{eq}$ and $F_{eq}$ with $\mu = 0$ and $T$,
determined from $\epsilon_0$, i.e.,
\be\label{eq:Tover}
T= \sqrt{\f{2}{\pi}}\frac{ (15\epsilon_0)^{1/4} }{\left( 8 N_c C_F+7 N_c N_f\right)^{1/4} }.
\ee
The excess gluons 
form a BEC. The total number of partons with $p>0$, $n_{eq}$, can be calculated from
\Eq{eq:n}, and the number density of the condensed gluons is given by
\be
N^0 = n_0 - n_{eq}.
\ee
  
\begin{figure}
\begin{center}
\includegraphics[width=0.45\textwidth]{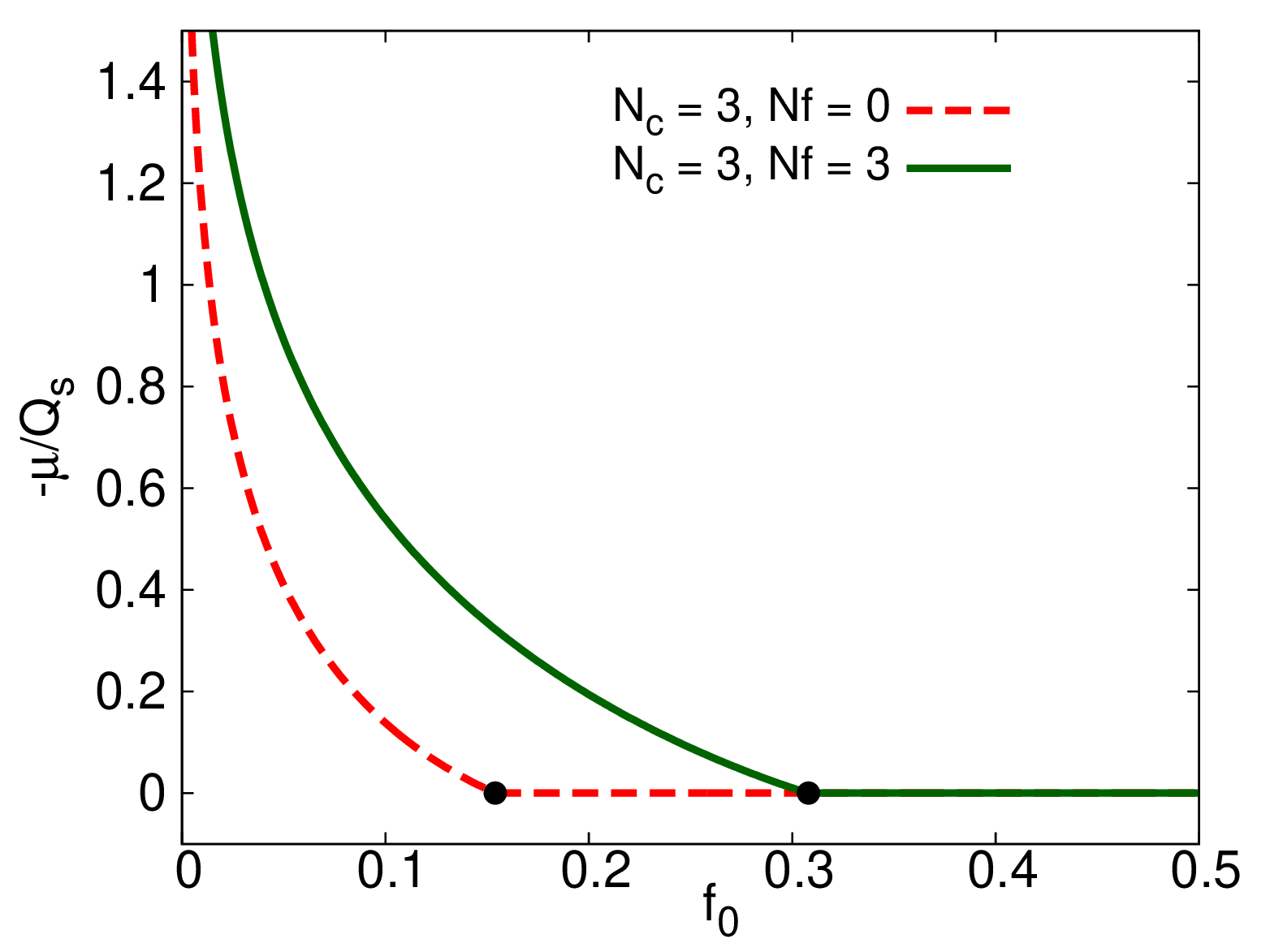}
\includegraphics[width=0.45\textwidth]{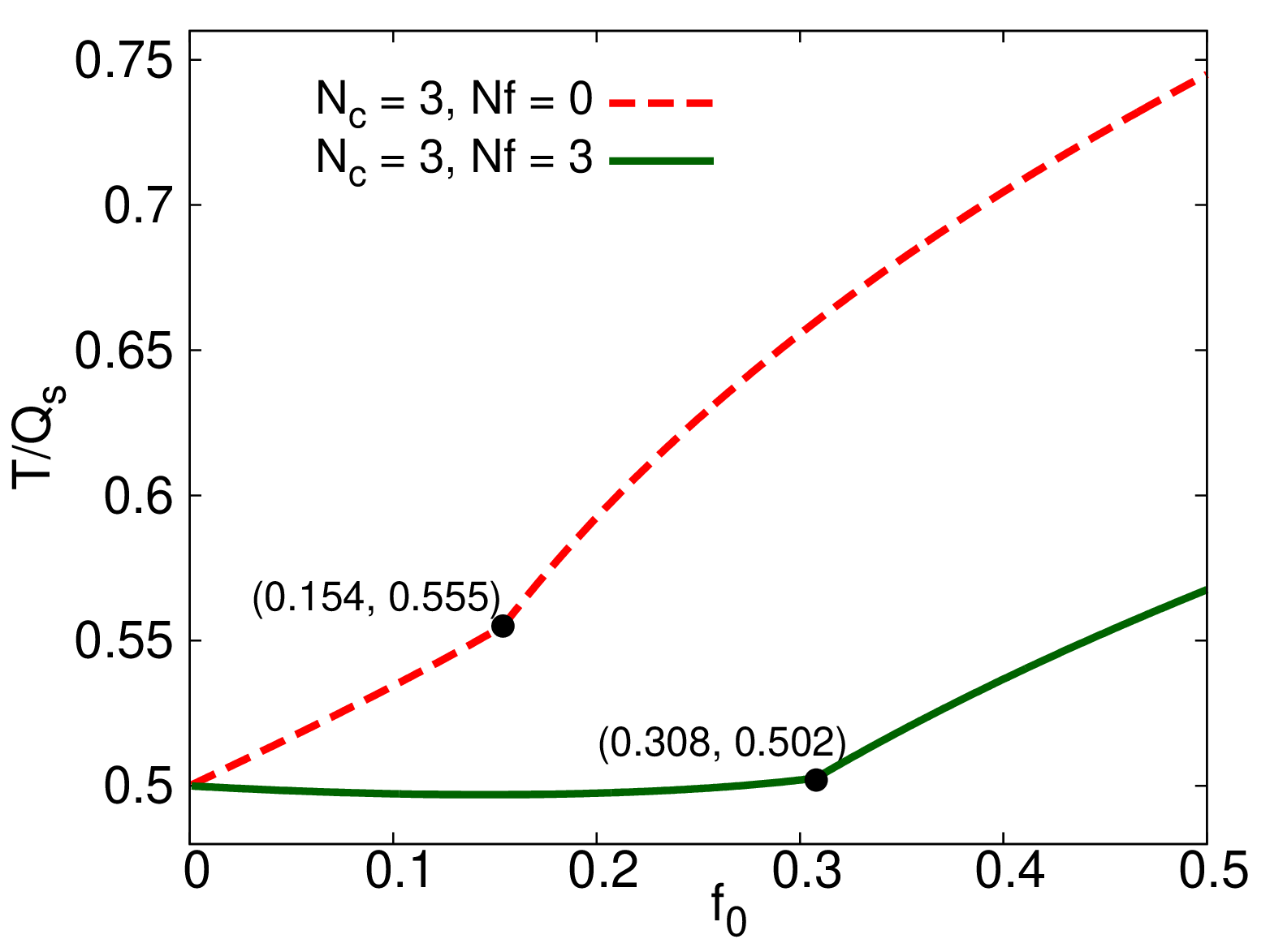}
\end{center}
\caption{(Color online)  The equilibrium temperature $T$ and chemical potential $\mu$ as a function of $f_0$. In both figures, the transition (marked by a black dot) from 
under-population to over-population occurs at $f_{0c} = 0.308$ for $N_f = 3$ (solid line) and $f_{0c} = 0.154$ for $N_f = 0$ (dashed line). 
For $f>f_{0c}$, the system is expected to be in a thermal equilibrium with vanishing $\mu$ and the excess gluons form a Bose condensate.
 }\label{fig:Tandmu}
\end{figure}
  
Let us take for example the system with the total energy and the particle number density
\bea
\epsilon_0 = \frac{f_0}{2\pi^2} N_c C_F Q_s^4, \qquad n_0 = \frac{f_0}{3\pi^2}  2 N_c C_F Q_s^3,\label{eq:n0}
\ea
as obtained from an initial distribution inspired by the color glass picture \cite{Blaizot:2011xf} (CGC)
\be\label{eq:f0}
f(0, p) = f_0~\theta\left(1 -\frac{p}{Q_s}\right), \qquad F(0, p) = 0
\ee
with $f_0>0$. The resulting dependence of $\mu$ and $T$ on $f_0$ is shown in Fig. \ref{fig:Tandmu}.  
The transition from under- to over-population happens at
\bea
&&f_{0c} =\frac{273375 (4 C_F+3 N_f)^4 \zeta(3)^4}{2 C_F (8 C_F+7 N_f)^3 \pi ^{12}}\simeq \frac{0.309 (4 C_F+3 N_f)^4}{C_F (8 C_F+7 N_f)^3},\label{eq:f0c}\\
&&T_c=\frac{45 \zeta (3) \left(4 C_F+3N_f\right)}{\pi ^4 \left(8C_F+7N_f\right)} Q_s \simeq \frac{0.555 \left(4 C_F+3N_f\right)}{\left(8C_F+7N_f\right)} Q_s.
\ea
Because the production of quarks and antiquarks effectively decreases the number of gluons, larger values of $f_{0c}$ are needed for $N_f>0$ than for $N_f=0$.
For example, $f_{0c} =0.308$ for $N_c = 3$ and $N_f = 3$, while $f_{0c} =0.154$ for $N_c = 3$ and $N_f = 0$. 
For $f<f_{0c}$ the system is under-populated. In this case $\mu$ and $T$ can be solved according to eq. (\ref{eq:muTunder}). 
For $f_0>f_{0c}$, the system becomes over-populated. The temperature is then given by eq. (\ref{eq:Tover}), that is,
\be
T = \frac{1}{\pi}\left( \frac{30 C_F f_0}{8 C_F +7 N_f}\right )^{\frac{1}{4}} Q_s,\qquad \mu=0.\label{eq:Teq}
\ee
\section{Thermalization of the quark-gluon plasma}\label{sec:results}
In this section we study the thermalization of a quark-gluon system whose initial distribution contains only gluons and is of the form given by eq. (\ref{eq:f0}). 
As discussed in the previous section, a BEC is expected to be formed 
when $f_0 >f_{0c}$ while when $f_0<f_{0c}$ there is no BEC in the equilibrium state. 
However, we shall show that even in the case $f_0 <f_{0c}$, when quarks are present, a BEC may appear for a short period of time due to the transient over-population of low momentum gluons. 
This occurs for $f_0>f_{0t}$, where $f_{0t}$ lies in the overlapping region between $\left.f_{0c}\right|_{N_f=0}$ 
and $\left.f_{0c}\right|_{N_f>0}$. In the following we study three different 
patterns of thermalization, each characterized by a specific value of $f_0$. In most cases, we take $N_f = 3$, in which case $f_{0t} \simeq 0.25937 < f_{0c} = 0.308$.

\subsection{Thermalization with BEC: $f_0 > f_{0c}$}

In Ref. \cite{Blaizot:2013:BEC}, it is shown that the onset of BEC in a dense system of gluons occurs in a finite time $\tau_c$.
For the initial condition eq. (\ref{eq:f0}), the transition value of $f_0$ from 
under-population to over-population is $\left. f_{0c}\right|_{N_f = 0} = 0.154$, which coincides with the value extracted from eq. (\ref{eq:f0c}) for $N_f=0$. 
In this subsection, we consider the effects of the quark production on the onset of BEC and manage to 
follow the evolution of the system, albeit very approximately, beyond $\tau_c$. The equilibration process is qualitatively the same for 
all the over-populated systems with the initial conditions (\ref{eq:f0}). We choose $f_0 = 0.4$ and $N_f = 3$ as a specific example to 
show the details of how the system evolves into a thermal equilibrium state with BEC.

\begin{figure}
\begin{center}
\includegraphics[width=0.45\textwidth]{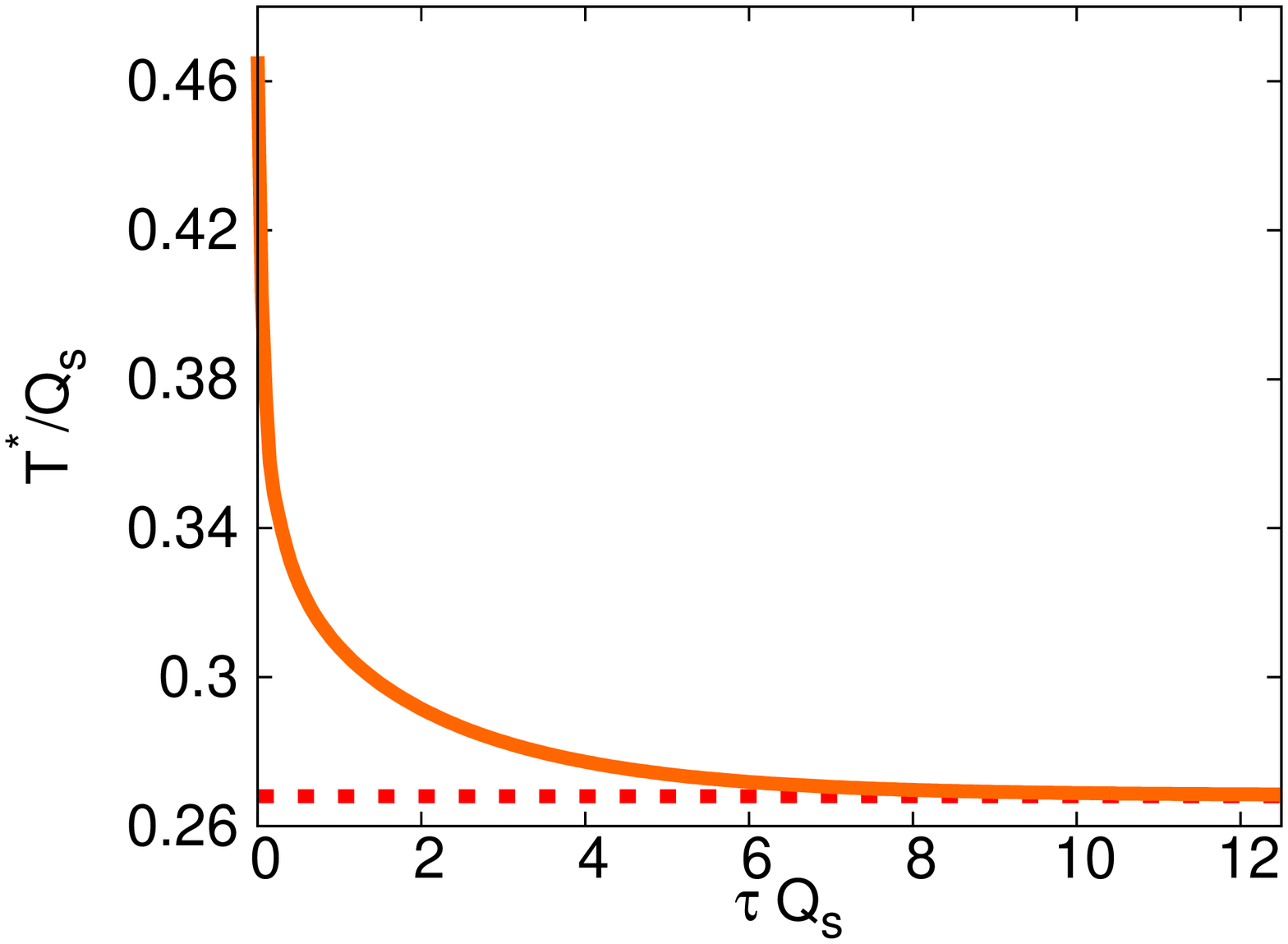}\hspace{0.05\textwidth}
\includegraphics[width=0.45\textwidth]{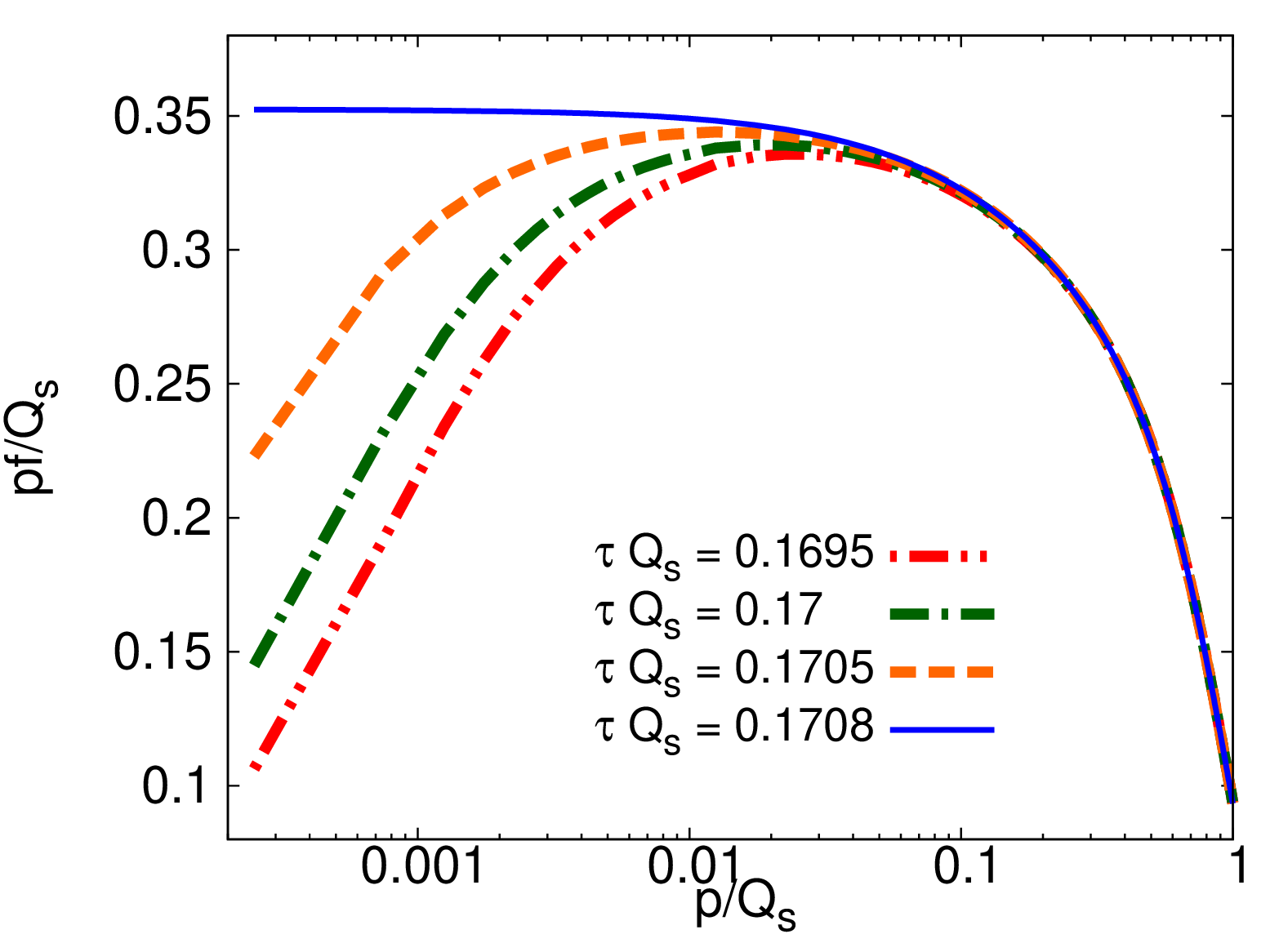}
\end{center}
\caption{(Color online)  The determination of $\tau_c$. The left panel shows $T^*$ as a function of $\tau$, which keeps decreasing to eventually approach the thermal equilibrium temperature $T=0.268~Q_s$.  
The right panel shows $p f$ near $\tau_c$, which is determined by eq. (\ref{eq:onset}). When $p\lesssim 0.1~Q_s$, the dashed curves are indistinguishable from the classical thermal distribution eq. (\ref{eq:onset1}), with a suitably adjusted $\mu^*$.  Here, $f_0 = 0.4$ and $N_f=3$, and $\tau_c~Q_s = 0.1708$.}\label{fig:taucdef}
\end{figure}

\begin{figure}
\begin{center}
\includegraphics[width=0.45\textwidth]{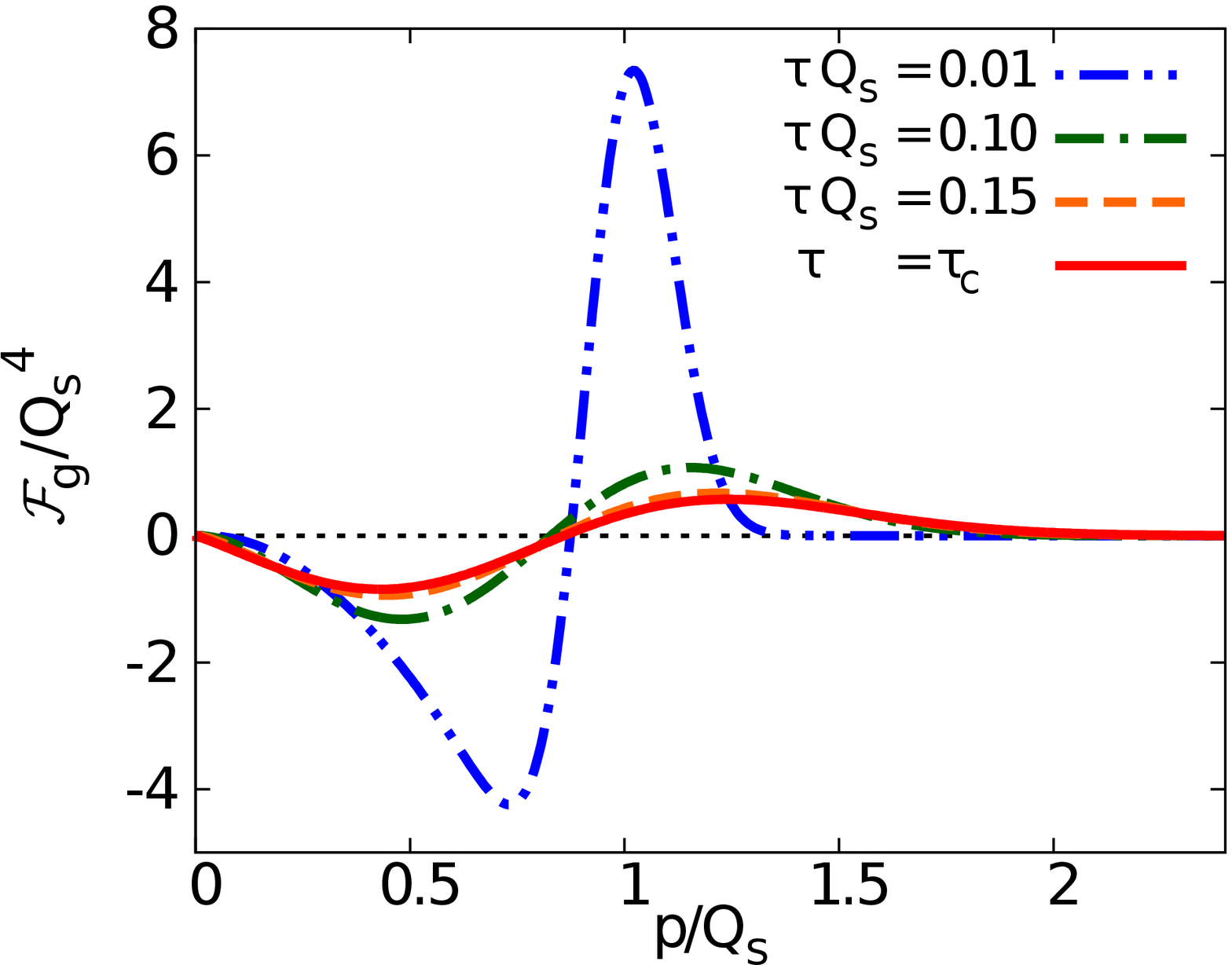}\hspace{0.05\textwidth}
\includegraphics[width=0.45\textwidth]{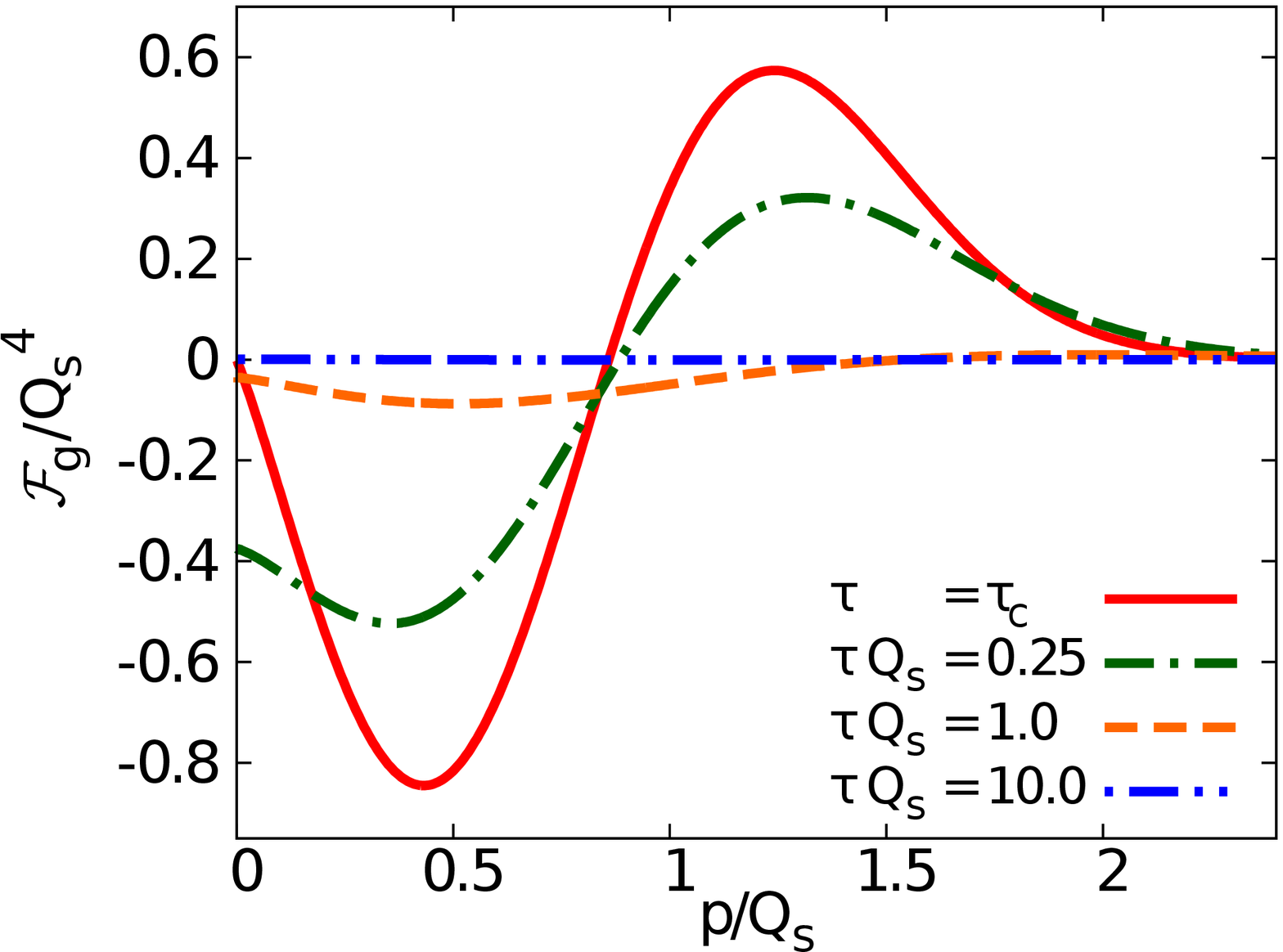}
\end{center}
\caption{(Color online)  The onset of gluon BEC. The gluon flux $\mathcal{F}_g$ at different times is shown as a function of $p$. Before $\tau_c \simeq 0.1708~Q_s^{-1}$, 
$\left.\mathcal{F}_g\right|_{p=0}$ vanishes (left panel). Right after $\tau_c$, $\left.\mathcal{F}_g\right|_{p=0}$ becomes finite and negative (right  panel). Here, $f_0 = 0.4$ and $N_f=3$.}\label{fig:jfover}
\end{figure}

\begin{figure}
\begin{center}
\includegraphics[width=0.45\textwidth]{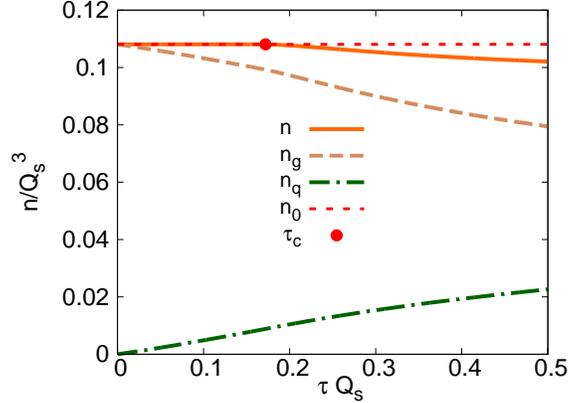}
\end{center}
\caption{(Color online)  Time evolution 
of the parton numbers around $\tau_c$. Before $\tau_c$, the total number of partons (with $p>0$) $n$ is equal to $n_0$. Right after $\tau_c$, it decreases and the difference between $n$ and $n_0$ is equal to the number density $N_0$ of gluons stored in the condensate. Here, $f_0 = 0.4$ and $N_f=3$.}\label{fig:over}
\end{figure}

The formation of BEC starts at a finite time $\tau = \tau_c$ when $f$ builds up the $1/p$ tail at small $p$ with the 
coefficient $c_{-1}$ given by \cite{Blaizot:2013:BEC}
\be\label{eq:onset}
c_{-1}  =  \f{I_b}{I_a}= T^*.
\ee
As discussed in Appendix \ref{app:series}, this is easily understood from the fact that the distribution function at small momentum is accurately described by the classical distribution function,
\be\label{eq:onset1}
f\simeq \f{T^*}{p-\mu^*_g},
\ee
with $T^*$ and $\mu^*_g$ time dependent parameters. 
The onset of BEC corresponds to the vanishing of the effective chemical potential, $\mu^*_g\to 0$, at which point $f \sim \f{T^*}{p}$. In our numerical simulation, eq. (\ref{eq:onset}) is used to determine the value of $\tau_c$
\footnote{In our code, $\tau_c$ is calculated as the moment when $\left. p f\right|_{p = p_{min}} = T^* = \f{I_a}{I_b}$ with $p_{min}$ the smallest momentum.}.
The left panel of Fig. \ref{fig:taucdef} shows how the effective temperature $T^*$ keeps decreasing until it eventually  approaches  the  equilibrium temperature $T$. The curve is completely smooth and does not show any indication of the onset of BEC that occurs at $ \tau_c~Q_s = 0.1708$ (for $f_0 = 0.4$ and $N_f=3$). The right panel of Fig. \ref{fig:taucdef} shows the time evolution of the gluon distribution function near $\tau_c$, and the approach to the singular behavior, $f(p)\sim 1/p$. The dashed curved are well fitted by the classical distribution (\ref{eq:onset1}). Before $\tau_c$, $f$ and $F$ are both analytic near $p=0$ and the gluon flux $\mathcal{F}_g$ vanishes at $p = 0$ (see the left panel of Fig. \ref{fig:jfover}): 
there is no accumulation of gluons at $p=0$. At $\tau = \tau_c$, $f$ becomes singular at $p=0$ but the gluon flux  
$\left.\mathcal{F}_g\right|_{p=0}$ still vanishes according to eqs. (\ref{eq:onset}) and (\ref{eq:fbc}). As shown in Fig. 
\ref{fig:over}, beyond this moment, the low momentum gluons keep accumulating, as $c_{-1}$ becomes larger than $T^*$. 
Our numerical simulation shows that after $\tau_c$ no solutions with vanishing  $\left.\mathcal{F}_g\right|_{p=0}$ are allowed by 
the transport equations in (\ref{eq:ftoSolve}) and (\ref{eq:FtoSolve}). As discussed in Appendix \ref{app:series}, one can find solutions 
beyond $\tau_c$ by providing boundary conditions according to eq. (\ref{eq:fbc}) (or eq. (\ref{eq:bcnumerics})). 
We have used
such solutions to describe the evolution of the system after $\tau_c$. Although this procedure ignores important coupling between the condensate
and the non-condensate particles, which may alter the details of the dynamics and perhaps the thermalization time scale, 
it has the advantage of providing a continuous transition to the correct equilibrium state.
As shown in the right panel of Fig. \ref{fig:jfover} and Fig. \ref{fig:over}, $\left.\mathcal{F}_g\right|_{p=0}$ becomes negative right after $\tau_c$, and correspondingly $n$ starts to decreases. 
This reflects the formation of a BEC, with the number density of condensed gluon, $N^0$, increasing according to 
$\dot N^0=-\dot n$.

\begin{figure}
\begin{center}
\includegraphics[width=0.45\textwidth]{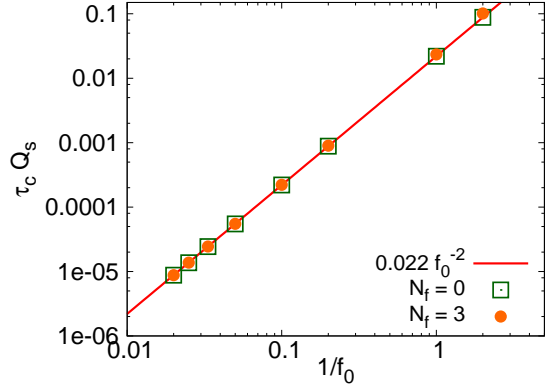}
\end{center}
\caption{(Color online)  $\tau_c$ as a function of $\f{1}{f_0}$. In both cases with $N_f = 0$ and $N_f = 3$, $\tau_c \propto \f{1}{f_0^2}$ 
and it is independent of $N_f$ for $f_0 \gtrsim 1$. }\label{fig:tauc}
\end{figure}

The dependence of $\tau_c$ on $f_0$ can be estimated parametrically at large $f_0$. Since $\tau_c$ decreases as $f_0$ 
increases\cite{Blaizot:2013:BEC}, one needs only study the time evolution of $f$ and $F$ at small $\tau$. This can be done by plugging
the linear expansions
\bea
f &\simeq& \bar f_0(p) + \tau \bar f_1(p),\nn
F &\simeq& \bar F_0(p) + \tau \bar F_1(p)=\tau \bar F_1(p),
\ea 
into eqs. (\ref{eq:ftoSolve}) and (\ref{eq:FtoSolve}) and keeping terms of $O(\tau^0)$. We obtain thus
\bea
&&\bar F_{1} = \frac{C_F^2 I_c {(0)}}{N_c p} \bar f_{0},\\
&&\bar f_{1}  = -\frac{1}{p^2} \left[ p^2 J_g(0, p) \right]' - \frac{N_f C_F I_c {(0)}}{N_c p}  \bar f_{0} \label{eq:ftoSolvetau},
\ea
where
\be
J_g{(0, p)}=
- N_c f_0^2 (1+f_0) Q_s^2\left[ 1+ \f{Q_s}{3} \delta(p-Q_s)  \right].
\ee 
In the limit $f_0\gg 1$ and $p\ll Q_s$, we have
\be
\bar f_{1}\sim \frac{2}{p} \left[ I_a {(0)} \bar f'_{0}+I_b{(0)} {\bar f_{0}}^2 \right]\sim f_0^3\f{Q_s^2}{p}.
\ee
Here, we have dropped the term $\propto \bar f'_{0} = f_0 \delta(p-Q_s)$, which vanishes at $p\ll Q_s$. Because the gluon flux vanishes 
at $p\sim Q_s$, $f$ does not change significantly at $p\sim Q_s$ and one has $Q_s f(Q_s)\sim Q_s~f_0$. Then $\tau_c$ can be estimated as 
the moment at which $p f$ at small $p$, $f^3_0 Q_s^2 \tau_c$, just becomes comparable with $p f$ at $p\sim Q_s$, $f_0Q_s$, which gives
\be\label{eq:taucPara}
\tau_c \sim \f{1}{f_0^2}\f{1}{Q_s}.
\ee
Since the quark production only contributes a term $\sim - N_f \f{f_0^2}{p}$ to $\bar f_{1}$, eq. (\ref{eq:taucPara}) is almost independent 
of $N_f$. Thus we do not expect quarks to affect the details of the transition to the BEC when $f_0$ is sufficiently large. The parametric behavior (\ref{eq:taucPara}) is confirmed by the numerical results shown in Fig. \ref{fig:tauc}, and it is actually valid for 
$f_0\gtrsim 1$. Therefore, we conclude that the formation of BEC starts at a time
\be\label{eq:tcpara}
t_c \sim \f{1}{(\alpha_s f_0)^2}\f{1}{Q_s}
\ee
for $f_0 \gtrsim 1$. Note however that for the specific value $f_0=0.4$ chosen for the numerical calculations presented in this subsection, $\tau_c=0.1708 \,Q_s^{-1}$, in slight deviation from this relation (which would yield $0.1375 \,Q_s^{-1}$).

\begin{figure}
\begin{center}
\includegraphics[width=0.45\textwidth]{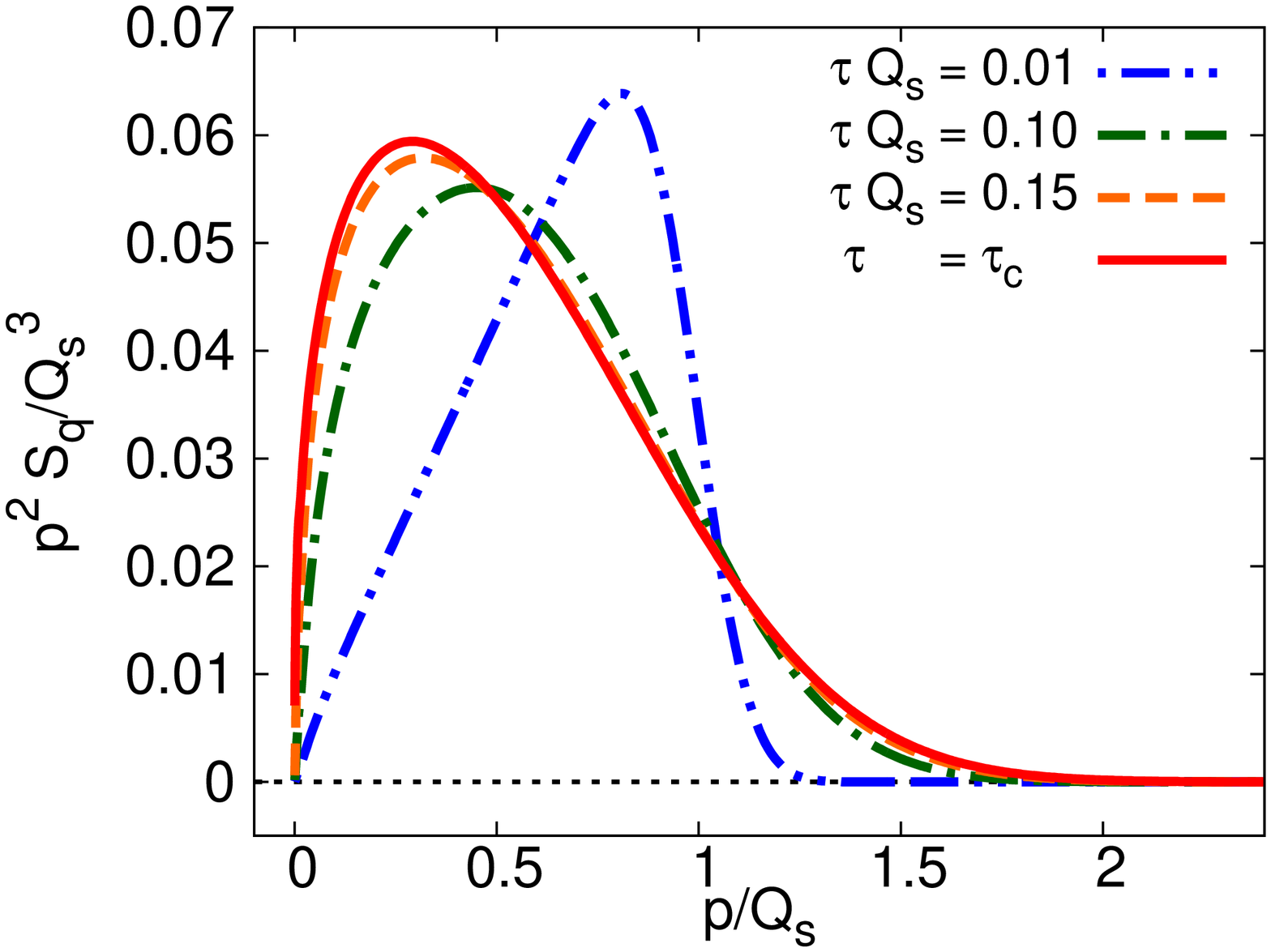}\hspace{0.05\textwidth}
\includegraphics[width=0.45\textwidth]{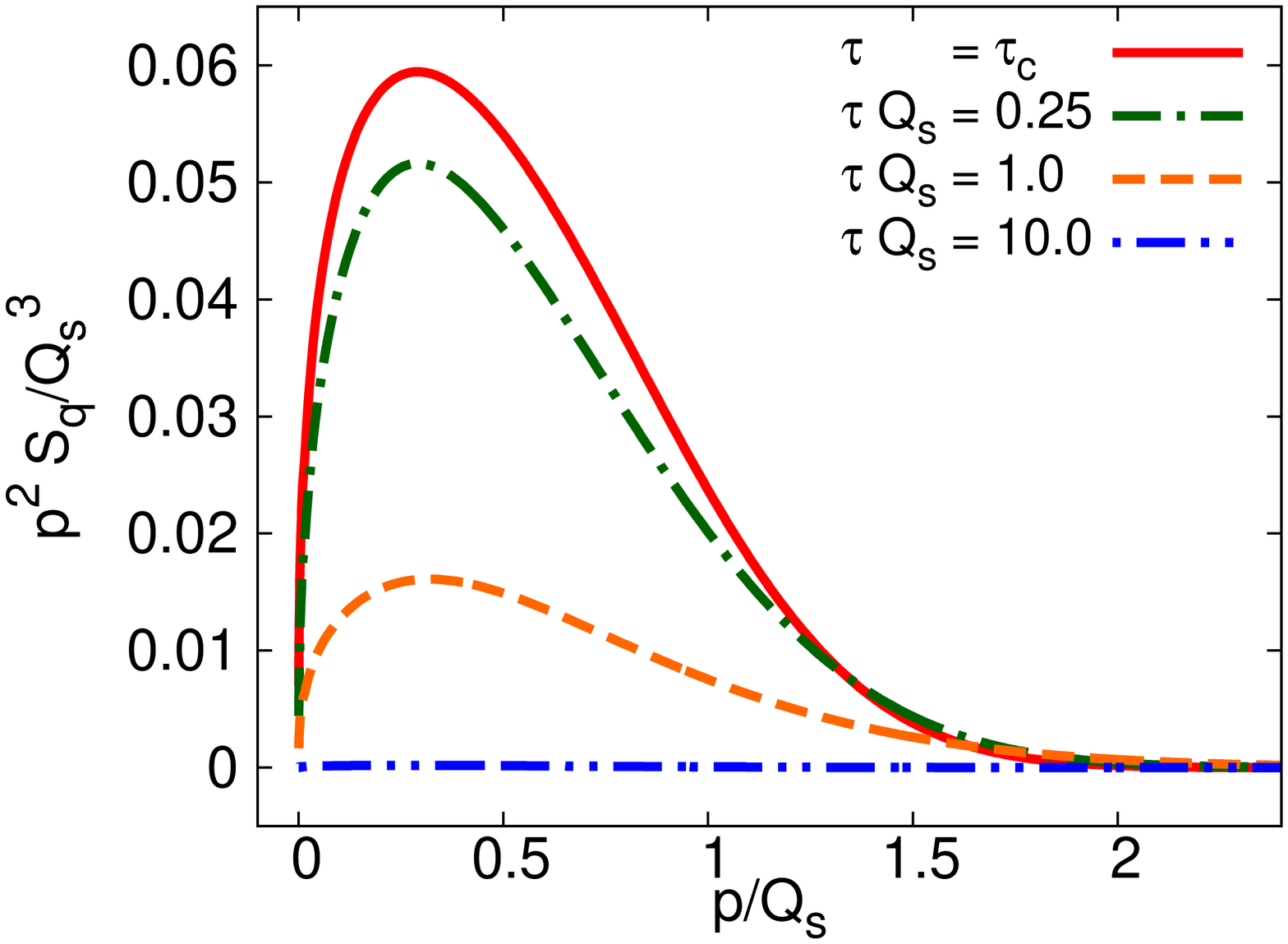}
\end{center}
\caption{(Color online)  Quark production. The production rate per unit momentum $p^2 S_q$ (source term multiplied by $p^2$) is shown as a function of $p$,  at different times before (left panel) and after (right panel)  the onset of BEC. Here, $f_0 = 0.4$ and $N_f=3$, in which case $\tau_c \simeq 0.1708~Q_s^{-1}$.}\label{fig:quarkProd}
\end{figure}

\begin{figure}
\begin{center}
\includegraphics[width=0.45\textwidth]{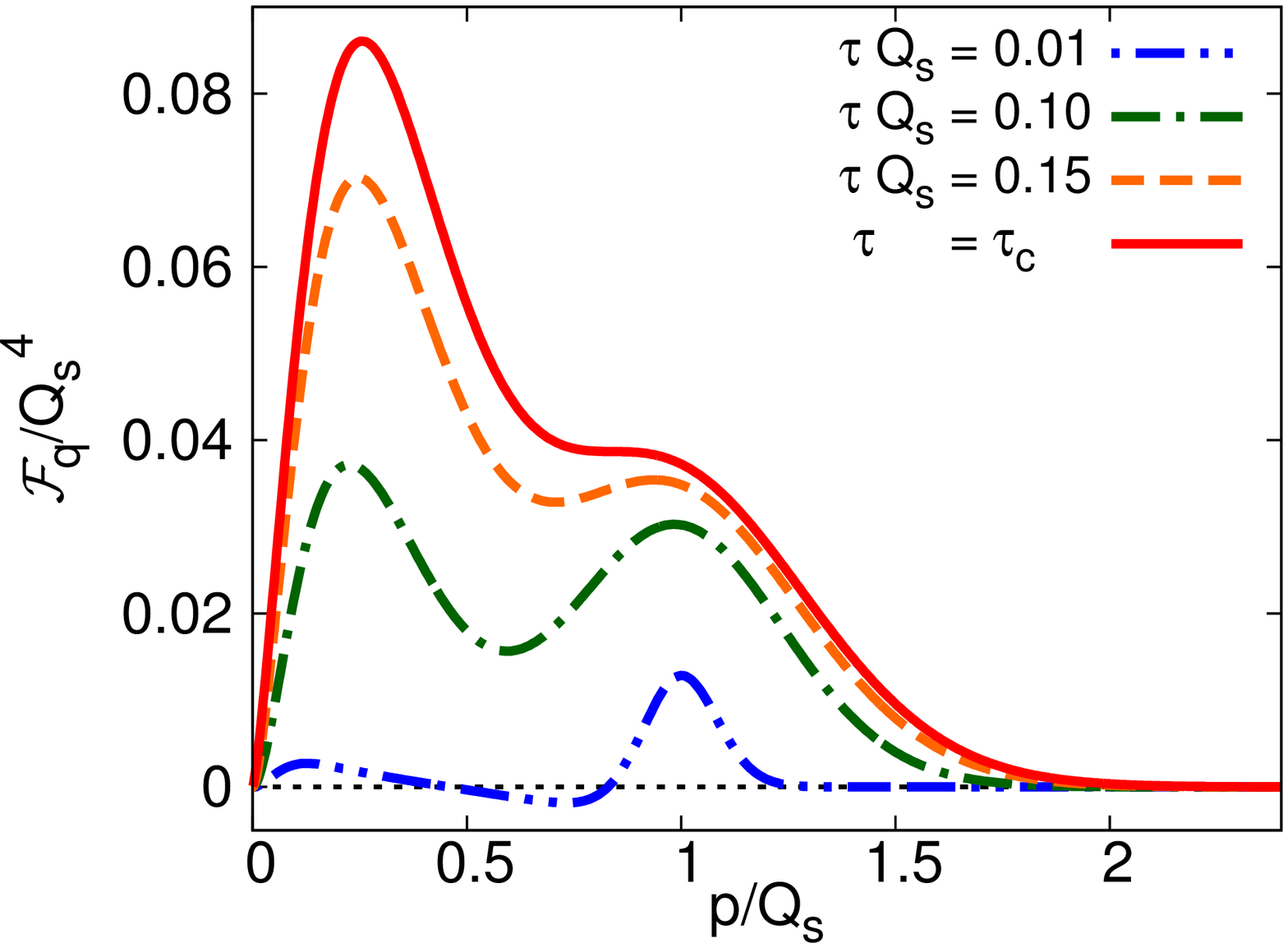}\hspace{0.05\textwidth}
\includegraphics[width=0.45\textwidth]{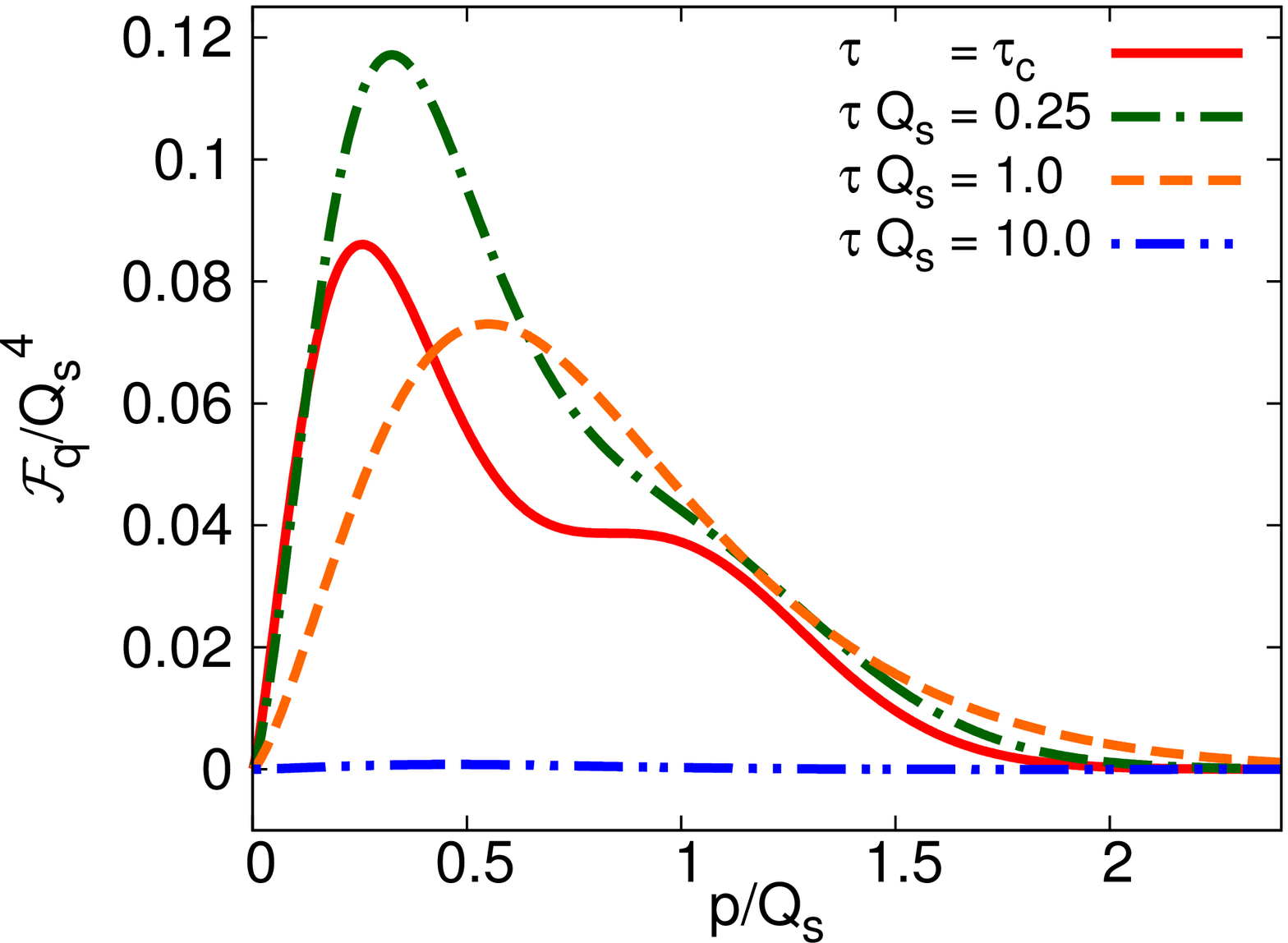}
\end{center}
\caption{(Color online)  The flux of the quark current as a function of momentum for different  times before (left panel) and after (right panel)  the onset of BEC.  Here, $f_0 = 0.4$ and $N_f=3$, in which case $\tau_c \simeq 0.1708~Q_s^{-1}$.}\label{fig:quarkProd2}
\end{figure}

We now consider the effect of quark production on the thermalization process. As we have already mentioned, inelastic processes involving quarks contribute both to the currents and to the source terms in eqs.~(\ref{boltz_equ}). At very early times, the gluon distribution function is approximately given by 
\be\label{eq:fdotpara}
\f{\partial}{\partial t} f \sim {\alpha_s^2} N_c^2 f_0^2 (1+f_0)\f{Q_s^2}{p} - {\alpha_s^2}N_f C_F f_0^2  \f{Q_s^2}{2 p}
\ee
for $p\ll Q_s$, and $p$ not too small. The first term on the right hand side of eq. (\ref{eq:fdotpara}) is due to the second part of the current (\ref{current_g}), the part proportional to the integral ${\cal I}_b$, that drives the increase of the population of soft gluons. The second term is 
due to the quark production. It acts in the opposite direction, thus hindering the growth of soft gluon modes. However, 
as shown in Fig.~\ref{fig:quarkProd}, after a short transient period of time, the quark production is peaked at small momenta. This is also confirmed by the plot of the quark flux plotted in Fig.~\ref{fig:quarkProd2}: the flux is the largest at small momenta, and continues to increase there all the way till the onset of BEC, and in some cases even beyond the BEC threshold, as revealed by the right hand side of Fig.~\ref{fig:quarkProd2}. In this regime, the quark production has no direct effect on the BEC itself. This is because, at small momenta, the outgoing quark current out of a small sphere of radius $p_0$  is compensated by the contribution to particle production in that small sphere  (i.e. by the source term, as can be verified explicitly by using the small $p$ expansions given in Appendix B, see in particular eq.~(\ref{currentssmallp}) showing that the constant contributions to the current are proportional to ${\cal I}_c$ and cancel with the source term, leaving a contribution linear in $p$). This leaves only the gluon current produced by elastic collisions as the source of variation of particle number in the small sphere. And indeed the gluon flux displayed in Fig.~\ref{fig:jfover} is very similar to that obtained for a purely gluonic system (see e.g. \cite{Blaizot:2013:BEC}). We have also verified that in the vicinity of the onset, the gluon chemical potential vanishes linearly with ($\tau_c-\tau$) within numerical accuracy, as it does in the purely gluonic system.

\begin{figure}
\begin{center}
\includegraphics[width=0.45\textwidth]{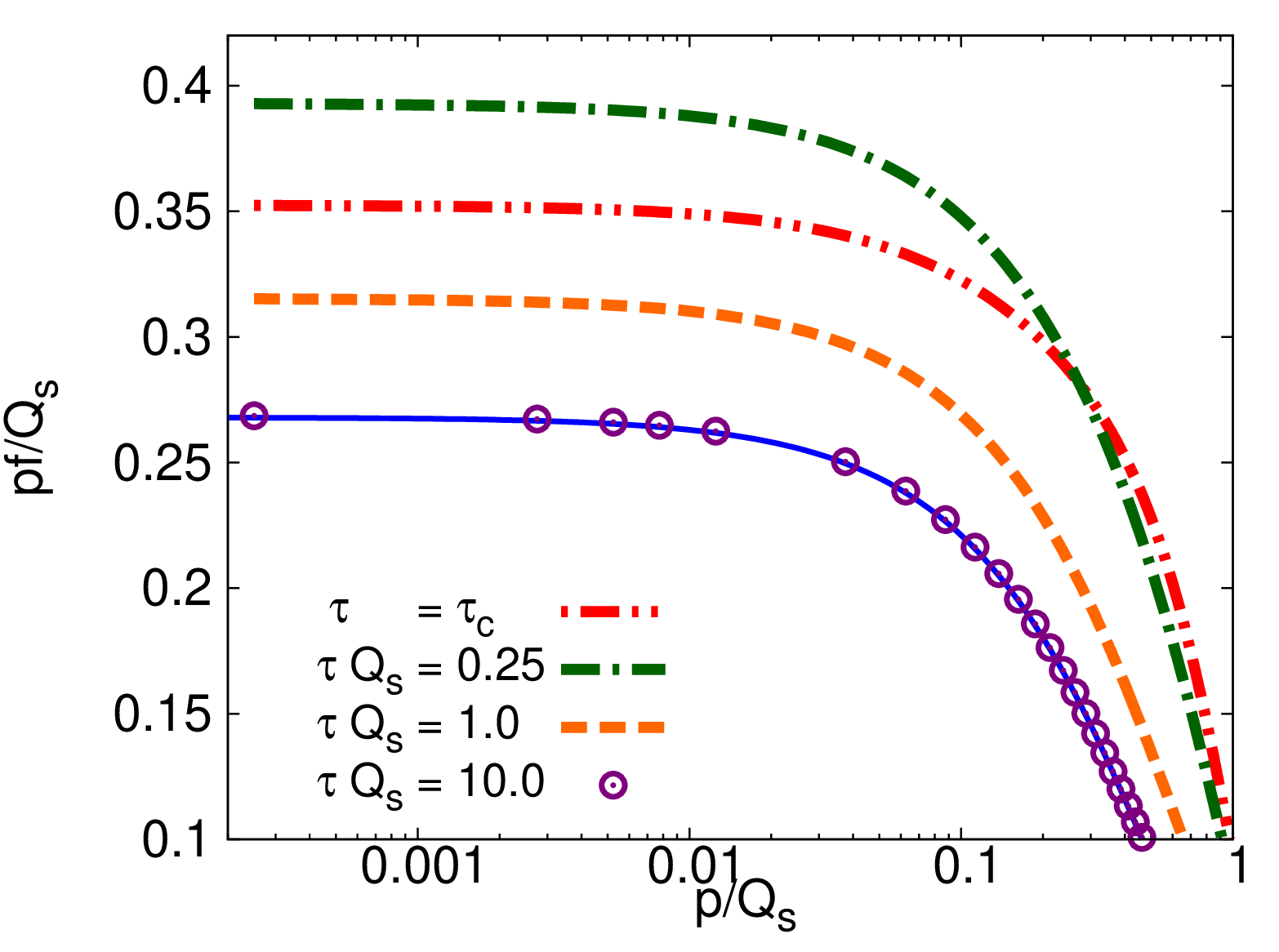}
\includegraphics[width=0.45\textwidth]{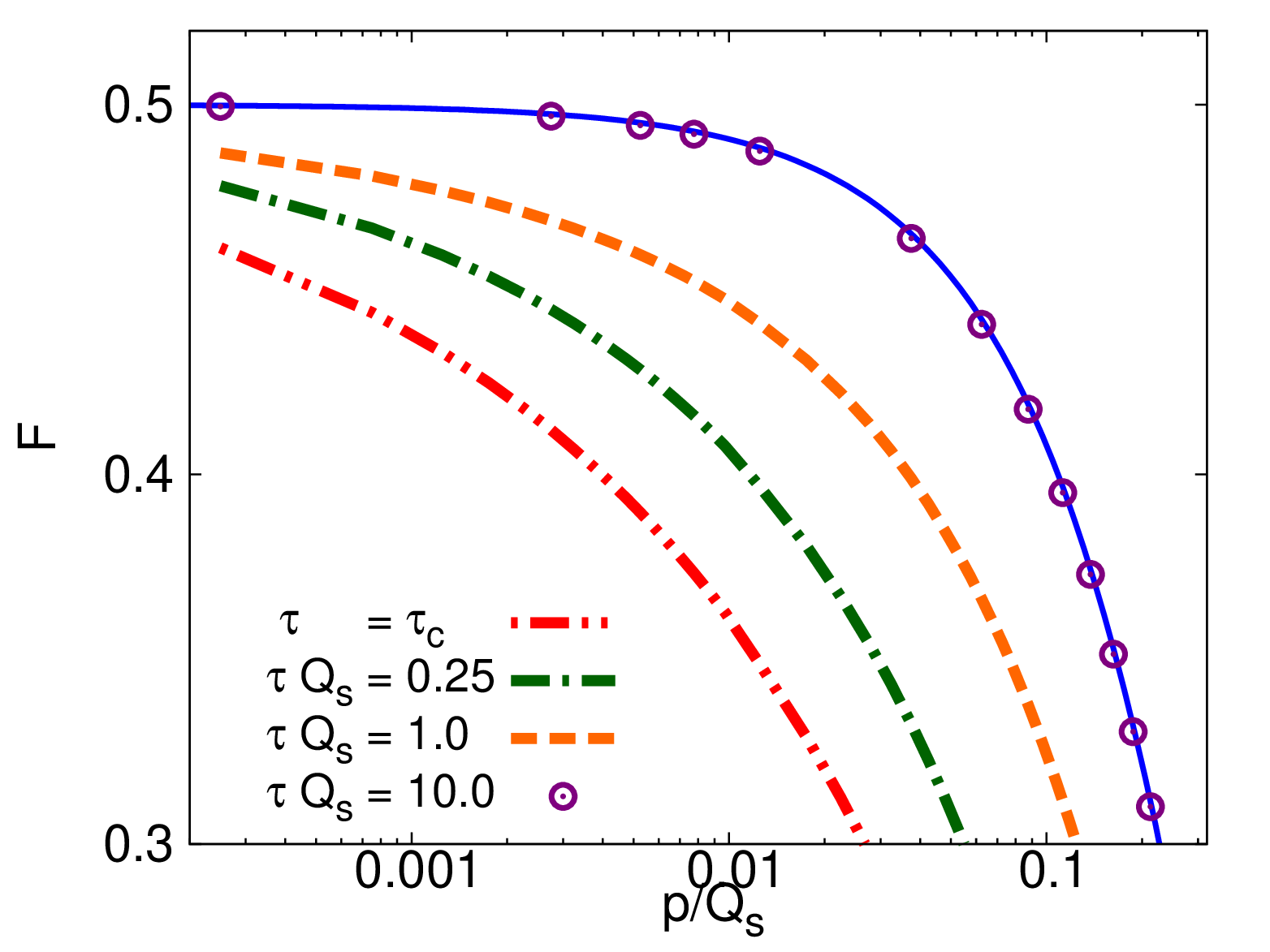}
\end{center}
\caption{(Color online)  Time evolution of $pf$ and $F$ after $\tau_c \simeq 0.1708~Q_s^{-1}$. During $ \tau_c \lesssim \tau \lesssim 0.25~Q_s^{-1}$ the gluon distribution $f$ increases at $p\lesssim Q_s$. After $\tau \simeq 0.25~Q_s^{-1}$, $f$ decreases to approach $f_{eq}$. In all the time $F$ increases. At $\tau\simeq 10~Q_s^{-1}$ $f$ and $F$ can be very well fitted by $f_{eq}$ and $F_{eq}$ with $\mu = 0$ and $T = 0.268~Q_s$ (solid blue lines). Here, $f_0 = 0.4$ and $N_f=3$.}\label{fig:beyond}
\end{figure}

\begin{figure}
\begin{center}
\includegraphics[width=0.45\textwidth]{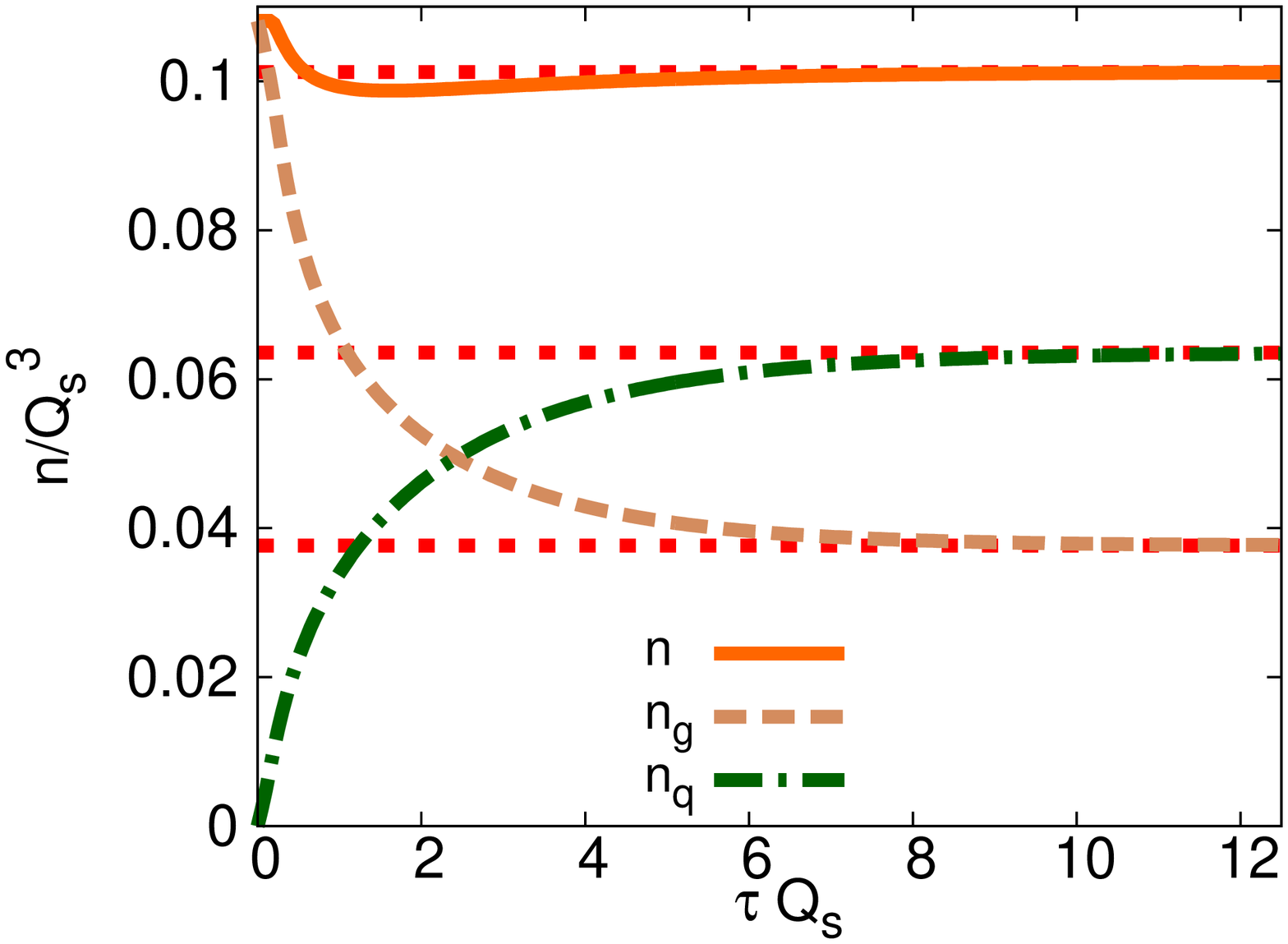}\hspace{0.05\textwidth}
\includegraphics[width=0.45\textwidth]{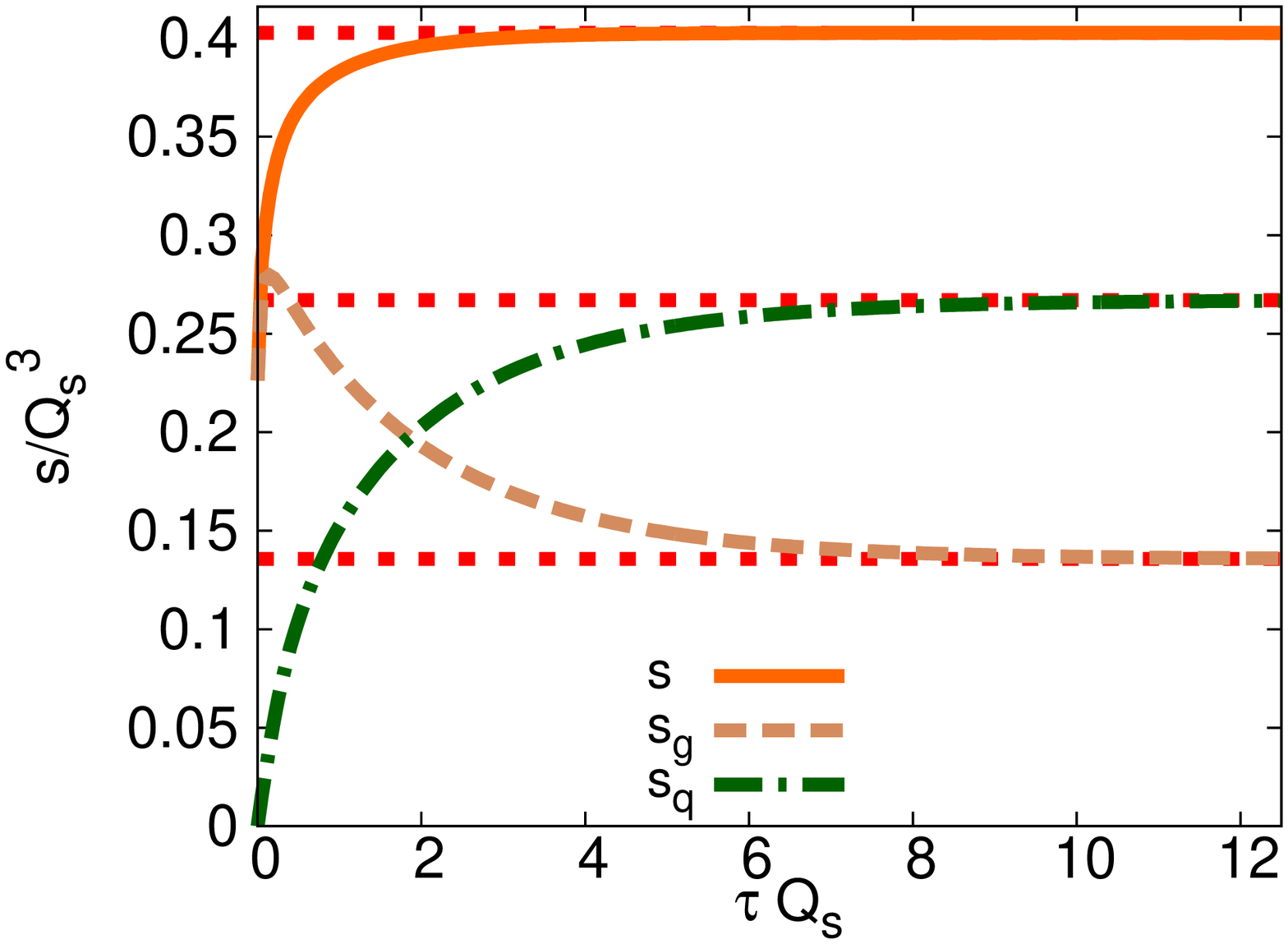}
\end{center}
\caption{(Color online)  Evolution of the number densities (left panel) and the entropy densities (right panel) of partons. The parameters $f_0 = 0.4$ and $N_f=3$ correspond to over-population. The formation of a BEC is seen in the left panel as the rapid decrease of the number density at small times (with a visible slight undershoot before reaching the equilibrium value).
The dotted red lines are the equilibrium values expected from thermodynamics. }\label{fig:f0040Nf3}
\end{figure}

For the chosen parameters, $N_f=3$, $f_0=0.4$, the thermal equilibrium can be only achieved after the formation of a gluon BEC. Fig. \ref{fig:beyond} shows how $f$ and $F$  evolve into 
thermal distributions after $\tau_c$. As we mentioned above, the number of low momentum gluons keeps growing right after $\tau_c$. 
This is a consequence of the relative small rate of quark production (see Fig. \ref{fig:quarkProd}) and condensate formation  
$\dot N^0$, in comparison with the growth rate of low momentum gluons due to the collisions. In the meantime, $\dot N^0$ increases 
because of the increase of $c_{-1}$ according to eq. (\ref{eq:np0t}). The occupation number of low momentum gluons stops growing and 
starts to decrease at a later time when the condensate formation rate and the quark production rate take over. 
Note that, as shown in \Fig{fig:quarkProd}, quark production takes place predominantly at low momentum. The high momentum quark modes
are populated by transport.
Afterwards, $f$ keeps 
decreasing while $F$ keeps increasing until the system achieves thermal equilibration. Fig. \ref{fig:f0040Nf3} shows the details about 
how the number and entropy densities evolve with $\tau$ and eventually reach the predicted values from thermodynamics in Sec. \ref{sec:equilibrium}.

Finally, let us discuss under which conditions the quark production from gluons can be neglected. First, as we have shown  
in the case $f_0\gtrsim1$, $\tau_c$ is (almost) independent of $N_f$. On the other hand, for $f_0\lesssim 1.0$ quark production delays 
the onset of BEC and $\tau_c$ increases as $N_f$ increases. For example, for $f_0 = 0.4$ $\tau_c \simeq 0.14~Q_s^{-1}$ with $N_f = 0$ and 
$\tau_c \simeq 0.1708~Q_s^{-1}$ with $N_f = 3$. This can be easily understood from eq. (\ref{eq:ftoSolvetau}): the production of quarks and 
antiquarks contributes a negative term $\propto -\f{N_f I_c{(0)}}{p} \bar f_{0}$ to $\bar f_{1}$, which obviously slows down the building-up 
of the $1/p$ tail of $f$ if $f_0$ is not large enough. Second, we observe that the quark production itself slows down the approach to 
thermalization . To make this statement more quantitative, we define an equilibration time $\tau_{eq}$ by the conditions
\be\label{eq:teqI}
\left| \f{T^*(\tau_{eq})}{T}-1 \right| \leq 0.05,~~\left| \f{n_g(\tau_{eq})}{n_{geq}}-1 \right| \leq 0.05,~~\left| \f{n_q(\tau_{eq})}{n_{qeq}}-1 \right| \leq 0.05,
\ee 
and
\be\label{eq:teqII}
\left| \f{s_g(\tau_{eq})}{s_{geq}}-1 \right| \leq 0.05,~~\left| \f{s_q(\tau_{eq})}{s_{qeq}}-1 \right| \leq 0.05,
\ee
where the values of the above quantities in thermal equilibrium are calculated using $f_{eq}$ and $F_{eq}$ with $T$ and $\mu$ given by eq. (\ref{eq:Teq}). For $f_0 = 0.4$, we find $\tau_{eq}\simeq 1.1~Q_s^{-1}$ with $N_f = 0$ and $\tau_{eq} \simeq 6.4~Q_s^{-1}$ with $N_f = 3$. And for $f_0 = 1.0$, we find $\tau_{eq}\simeq 0.86~Q_s^{-1}$ with $N_f = 0$ and $\tau_{eq} \simeq 4.8~Q_s^{-1}$ with $N_f = 3$. Thus, the presence of quarks increases the thermalization time by typically a factor of 5 (for $N_f = 3$) (we should keep in mind however that this estimate suffers from the uncertainties related to our very approximate description of the dynamics beyond the onset of BEC).

\subsection{Thermalization without BEC: $f_0 \leq f_{0t}$}
\begin{figure}
\begin{center}
\includegraphics[width=0.45\textwidth]{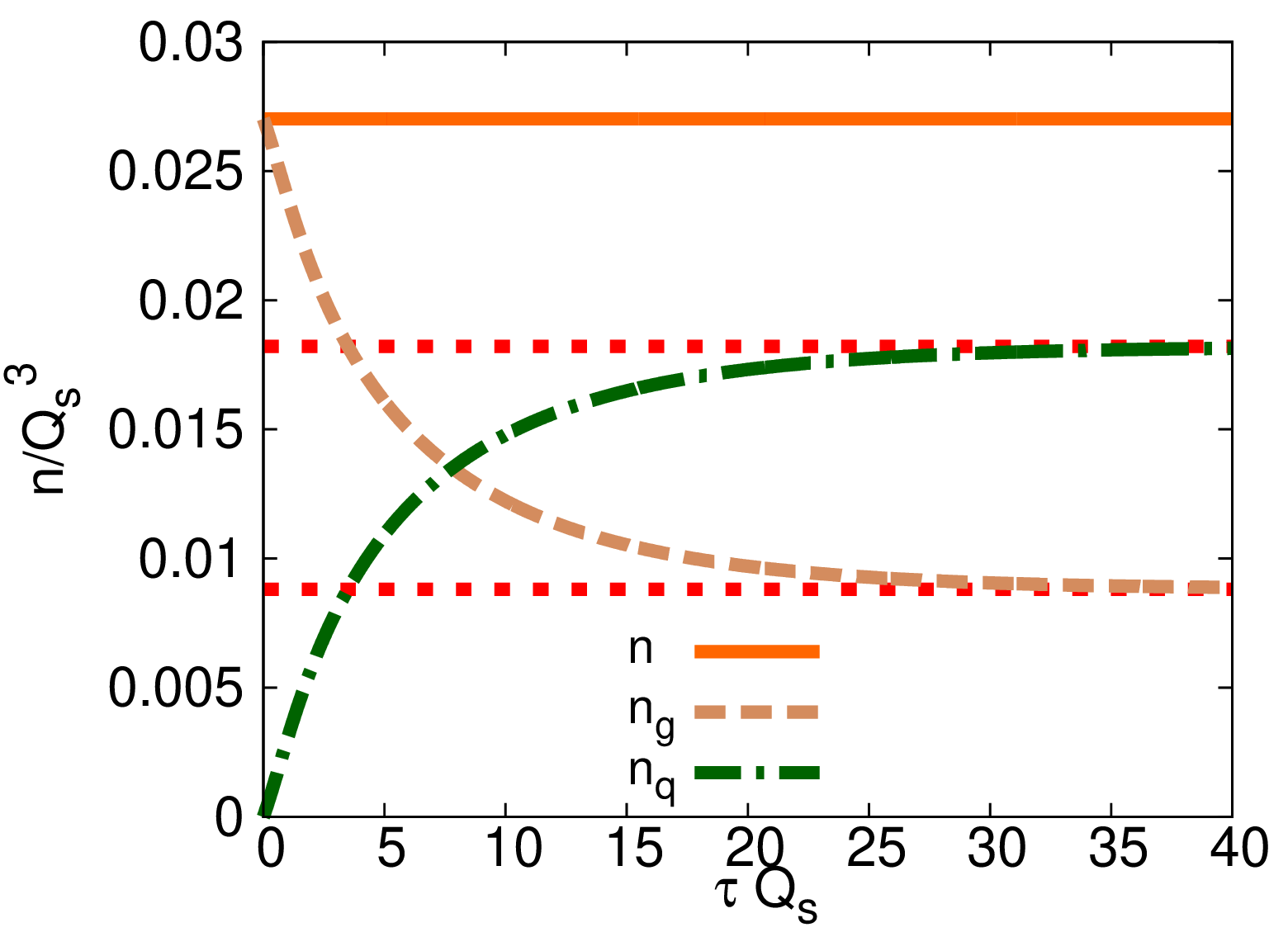}\hspace{0.05\textwidth}
\includegraphics[width=0.45\textwidth]{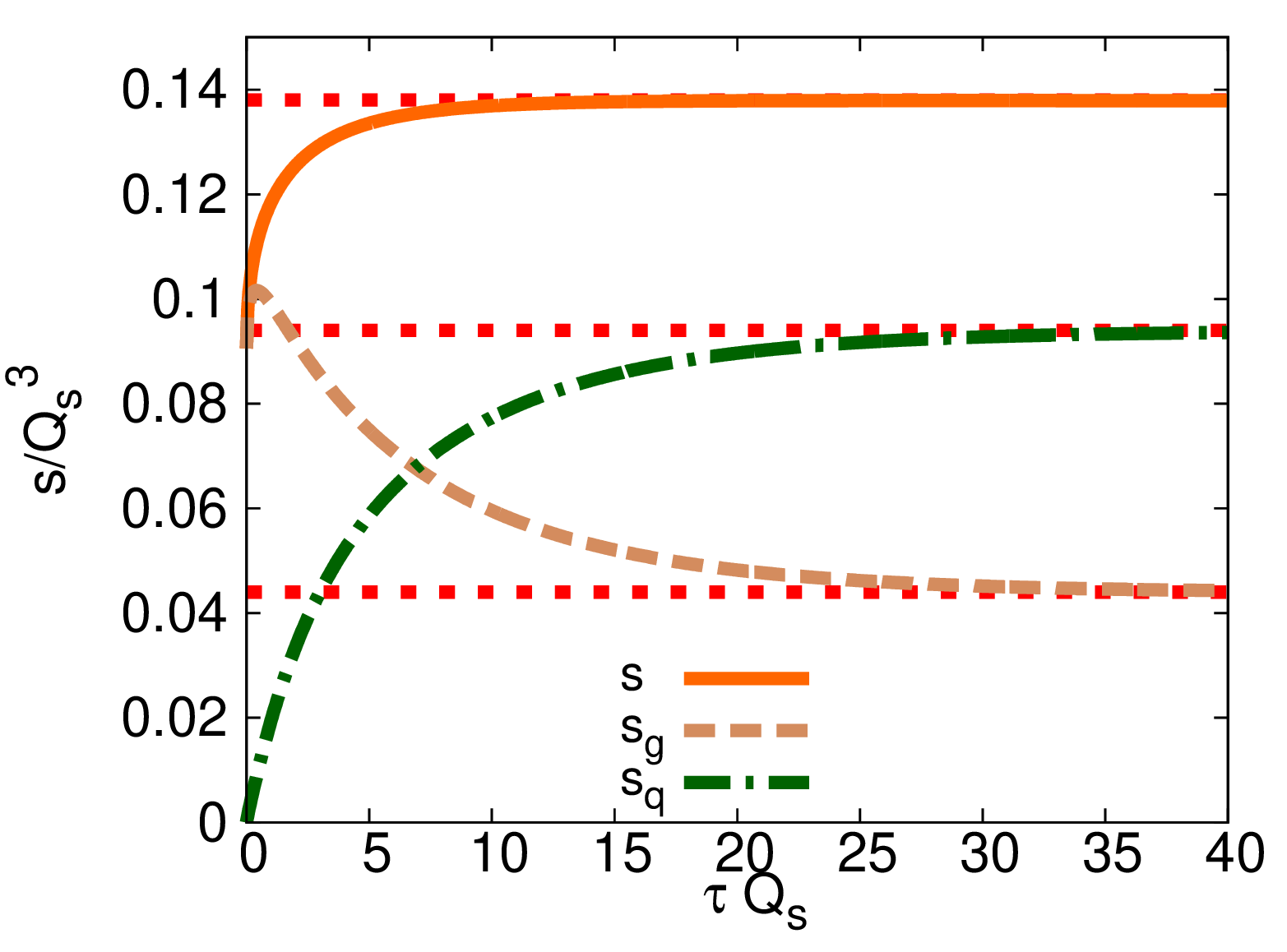}
\end{center}
\caption{(Color online)  Evolution of the number densities (left panel) and the entropy densities (right panel) of partons in the under-populated case ($f_0 = 0.1$ and $N_f = 3$). The horizontal dotted red lines are the equilibrium values expected from thermodynamics.}\label{fig:f0010Nf3}
\end{figure}

For the initial distribution (\ref{eq:f0}), with $f_0 \leq f_{0t}$, the quark-gluon system will achieve thermal equilibration without 
the formation of a BEC. Our numerical results verify that the thermal equilibrium temperature $T$ and the negative chemical potential 
$\mu$ are exactly those predicted by solving eq. (\ref{eq:muTunder}). In those cases, the features of the thermalization process are 
qualitatively the same for all $f_0$. The quarks and antiquarks are produced from the the process $gg\to q\bar{q}$, which causes the 
gluon number to decrease keeping the total parton number constant. The entropy density of gluons becomes smaller at later times but 
the total entropy density always increases. Fig. \ref{fig:f0010Nf3} shows the details about how the number and entropy densities of the 
system with $f_0 = 0.1$ and $N_f = 3$ evolve into their predicted values in thermal equilibrium.
These curves are quite similar to those in \Fig{fig:f0040Nf3}, with the noticeable difference that here the parton number $n$ is exactly conserved.

\begin{figure}
\begin{center}
\includegraphics[width=0.45\textwidth]{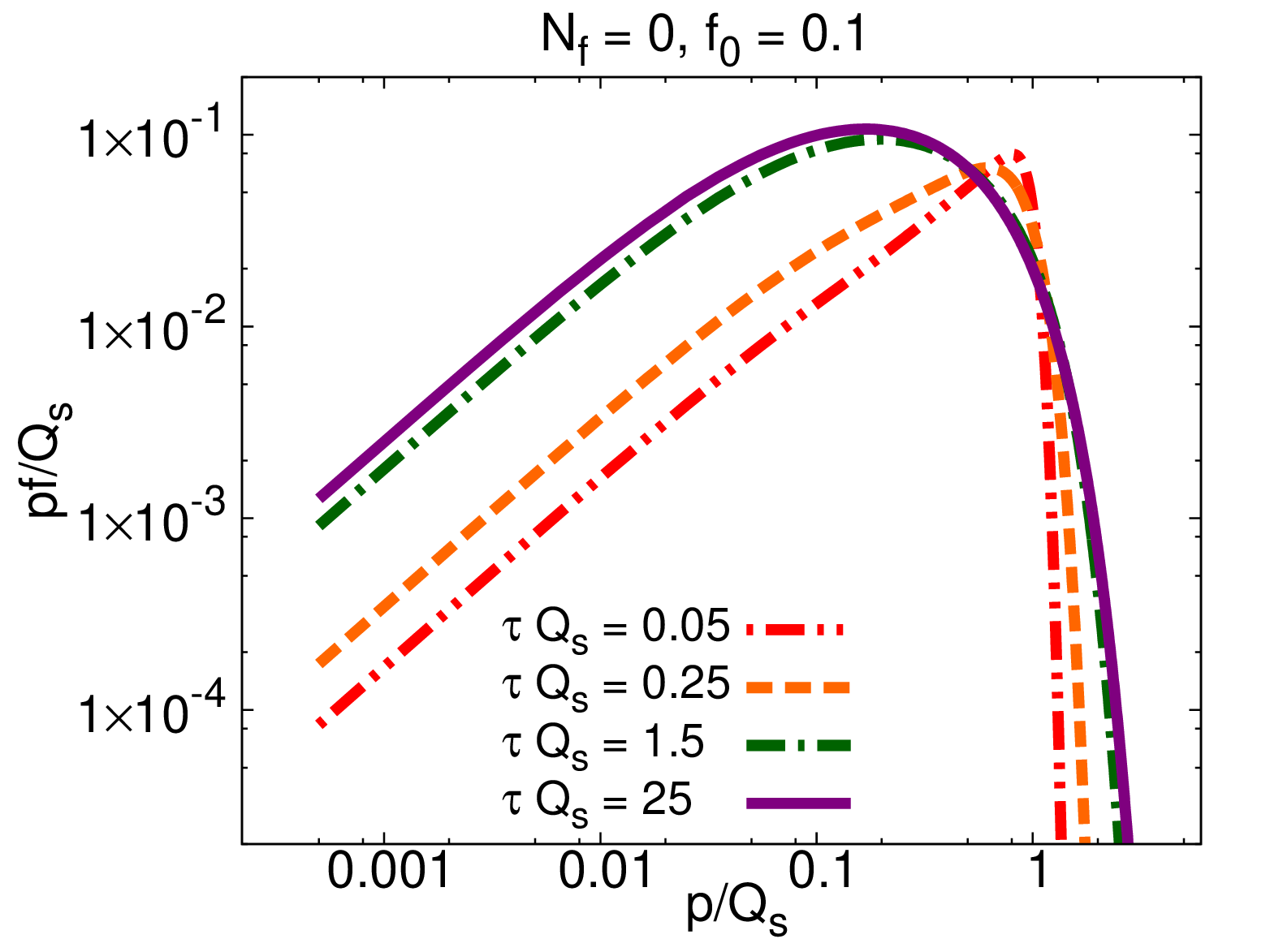}\hspace{0.05\textwidth}
\includegraphics[width=0.45\textwidth]{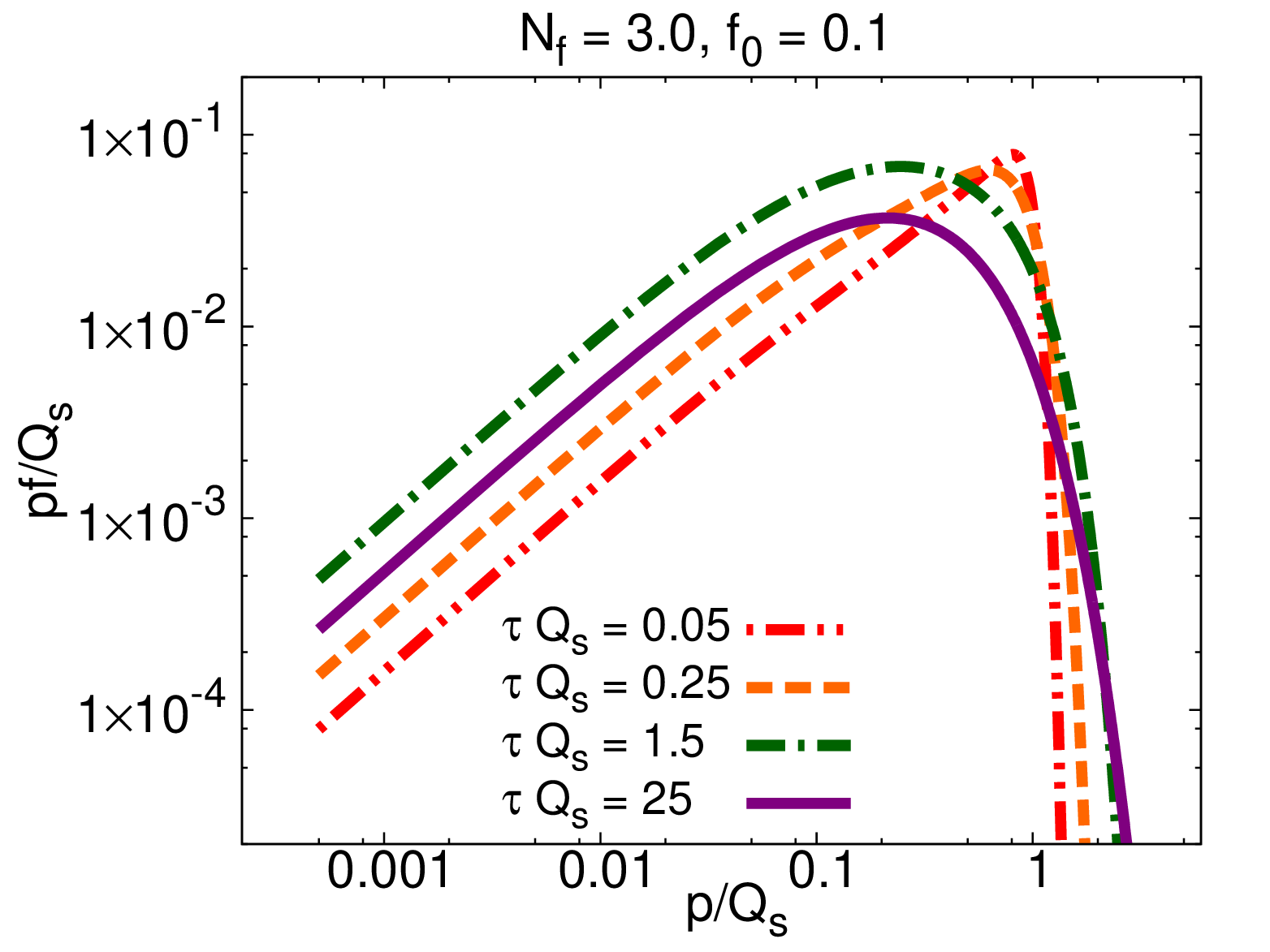}
\end{center}
\caption{(Color online)  \label{fig:pfNf} Evolution of the gluon distribution $f$ in the underpopulated case corresponding to $f_0=0.1$ and  $N_f = 0$ (left panel) and $N_f = 3$ (right panel). In both cases, the solid curves at $\tau~Q_s = 25$ are thermal 
equilibrium distributions. The left panel shows that for $N_f = 0$ the number of low momentum gluons continuously  increases until 
the system achieves thermal equilibrium. The right panel shows that for $N_f = 3$ the number of low momentum gluons overshoots that in the thermal distribution before the system eventually thermalizes.}
\end{figure}

With quark production turned off ($N_f=0$), the system of gluons with $f_0> \left.f_{0c}\right|_{N_f=0} =0.154$ thermalizes with the formation 
of BEC. As discussed in the previous subsection, the quark production contributes a term $\propto -N_f \f{f_0^2}{p}\tau$ to $f$ in the 
early time. For $f_{0t}>f_0> \left.f_{0c}\right|_{N_f=0} =0.154$, this term is large enough to prevent $f$ from building up the $1/p$ 
tail near $p=0$, 
thereby inhibiting the formation of a BEC.

For the same $f_0$, the system with $N_f \geq 3$ has a lower equilibrium $T$ and a smaller $\mu$ than that with $N_f = 0$, as 
shown in Fig. \ref{fig:Tandmu}. Such a difference causes the under-populated system to thermalize in a different pattern.  
An example with $f_0 = 0.1$ is shown in Fig. \ref{fig:pfNf}. For $N_f = 0$, the number of low momentum gluons continues 
to increase until the system achieves thermal equilibrium. For $N_f = 3$, the occupation number of gluons with $p\lesssim Q_s$ first 
reaches a maximum value which is higher than that in thermal equilibrium. Such an excess of gluons can not be tamed by the quark production 
until the late stages of equilibration.  Let us define two effective chemical potentials
\be
\mu^*_g \equiv -T^* \log\left(1+\f{1}{f(0)} \right),\qquad \mu^*_q \equiv -T^* \log\left(\f{1}{F(0)} - 1 \right),
\ee
which are both equal to $\mu$ after the system thermalizes.  If $f_0 < f_{0t}$, $f$ near $p=0$ can be approximated by $f_{eq}$ with $\mu_g^*/T^*<0$. 
Given $f_0$, one can determine the largest value of $\mu_g^*/T^*$ numerically. $f_{0t}$ is defined by the value of $f_0$ for which the 
largest value of $\mu_g^*/T^*$ is zero. At the moment when $\mu_g^*/T^*=0$, $f$ looks like $f_{eq}$ with a vanishing $\mu$ near $p=0$. 
In the next subsection, we shall show that BEC can be formed due to such a transient excess of gluons  in the system with $f_0 > f_{0t}$ 
and $N_f >0$. 

Like in the over-populated case, the quark production delays thermalization. The equilibration time $\tau_{eq}$ is redefined by replacing $n_g/n_{geq}$ and $n_q/n_{qeq}$ respectively by $\mu^*_g/\mu$ 
and $\mu^*_q/\mu$ in eq. (\ref{eq:teqI}). In this definition, the first condition in eq. (\ref{eq:teqI}) is a sufficient condition for $f$ 
and $F$ to be approximately equal to those in thermal equilibrium at small $p$ while the second condition in eq. (\ref{eq:teqII}) acts as 
a constraint to the shape of $f$ and $F$ for the full range of $p$. For $f_0 = 0.1$, we find $\tau_{eq} \simeq 5.5~Q_s^{-1}$ with $N_f = 0$ 
and $\tau_{eq}\simeq 25~Q_s^{-1}$ with $N_f = 3$ (again we observe that quark production delays the equilibration time by a factor of $\sim 5$).

\subsection{Thermalization with transient BEC: $f_{0c}>f_0 > f_{0t}$}

\begin{figure}
\begin{center}
\includegraphics[width=0.45\textwidth]{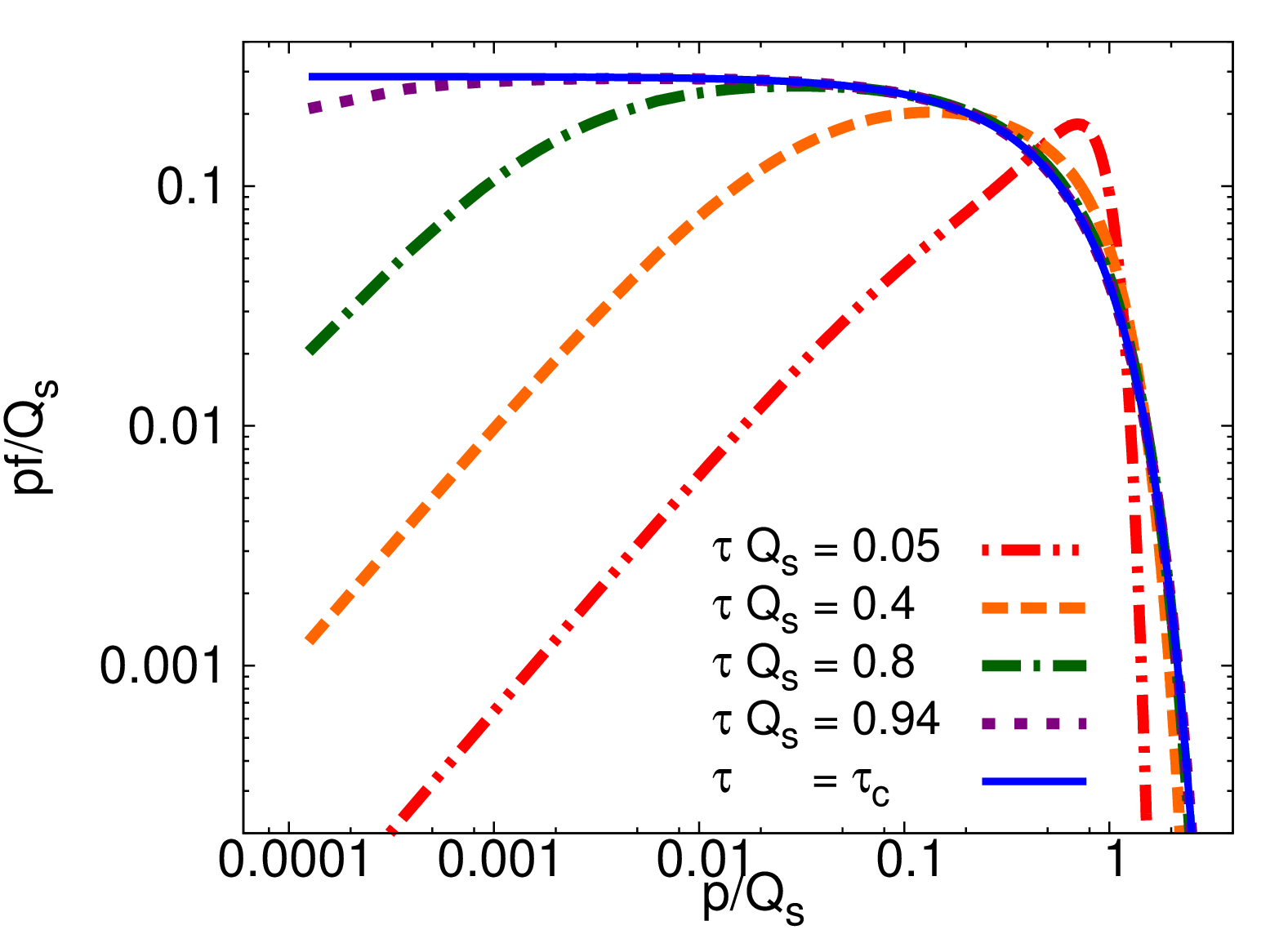}\hspace{0.05\textwidth}
\includegraphics[width=0.45\textwidth]{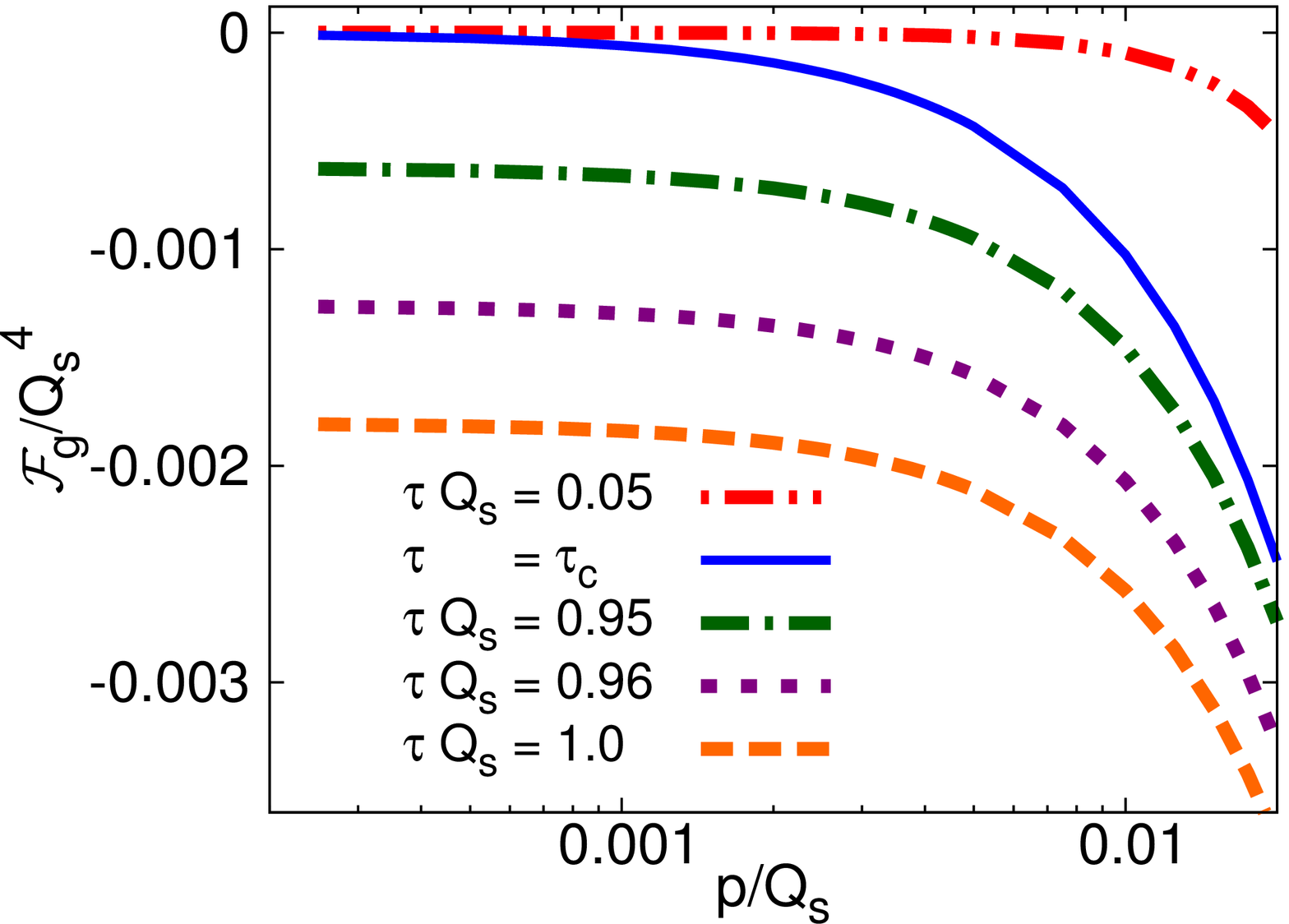}
\end{center}
\caption{(Color online)  Formation of transient BEC. The left panel shows the time evolution of $pf$ as a function of $p$. The $1/p$ tail of $f$ is built up at $\tau = \tau_c = 0.947~Q_s^{-1}$. The right panel shows the small $p$ behavior of the gluon flux $\mathcal{F}_g$ at different times. Right after $\tau_c$, it becomes negative. Here, $N_f =3$ and $f_{0t}<f_0 = 0.26 < f_{0c} = 0.308$.}\label{fig:tran}
\end{figure}

A transient BEC can be formed in the under-populated system with $f_0 > f_{0t}$. Let us choose the system with  $f_0 = 0.26$ and $N_f = 3$ 
as an example. As shown in the left panel of Fig. \ref{fig:tran}, $f$ starts to become singular at $p = 0$ at $\tau_c = 0.947~Q_s^{-1}$. 
At this moment,  the gluon flux $\mathcal{F}_g$ still vanishes at $p=0$ because $c_{-1} = T^*$ (see eq.~(\ref{eq:ndot})). However, $c_{-1}$ has a 
tendency to increase due to the further accumulation of small momentum gluons. Like in the over-populated case, the solution to the transport 
equations exists after $\tau_c$ only if boundary conditions with a non-vanishing $\left.\mathcal{F}_g\right|_{p=0}$ are provided. Using the 
boundary conditions in eq. (\ref{eq:bcnumerics}) to solve the transport equations, we are able to follow the subsequent evolution of the 
system. $\mathcal{F}_g$ is found to become negative at $p=0$ right after $\tau_c$, which is shown in  the right panel of Fig. \ref{fig:tran}. 
This negative gluon flux reflects the formation of a BEC, and the number density of condensed particles can be calculated from the gluon flux at $p=0$ according to 
eq. (\ref{eq:np0t}).

\begin{figure}
\begin{center}
\includegraphics[width=0.45\textwidth]{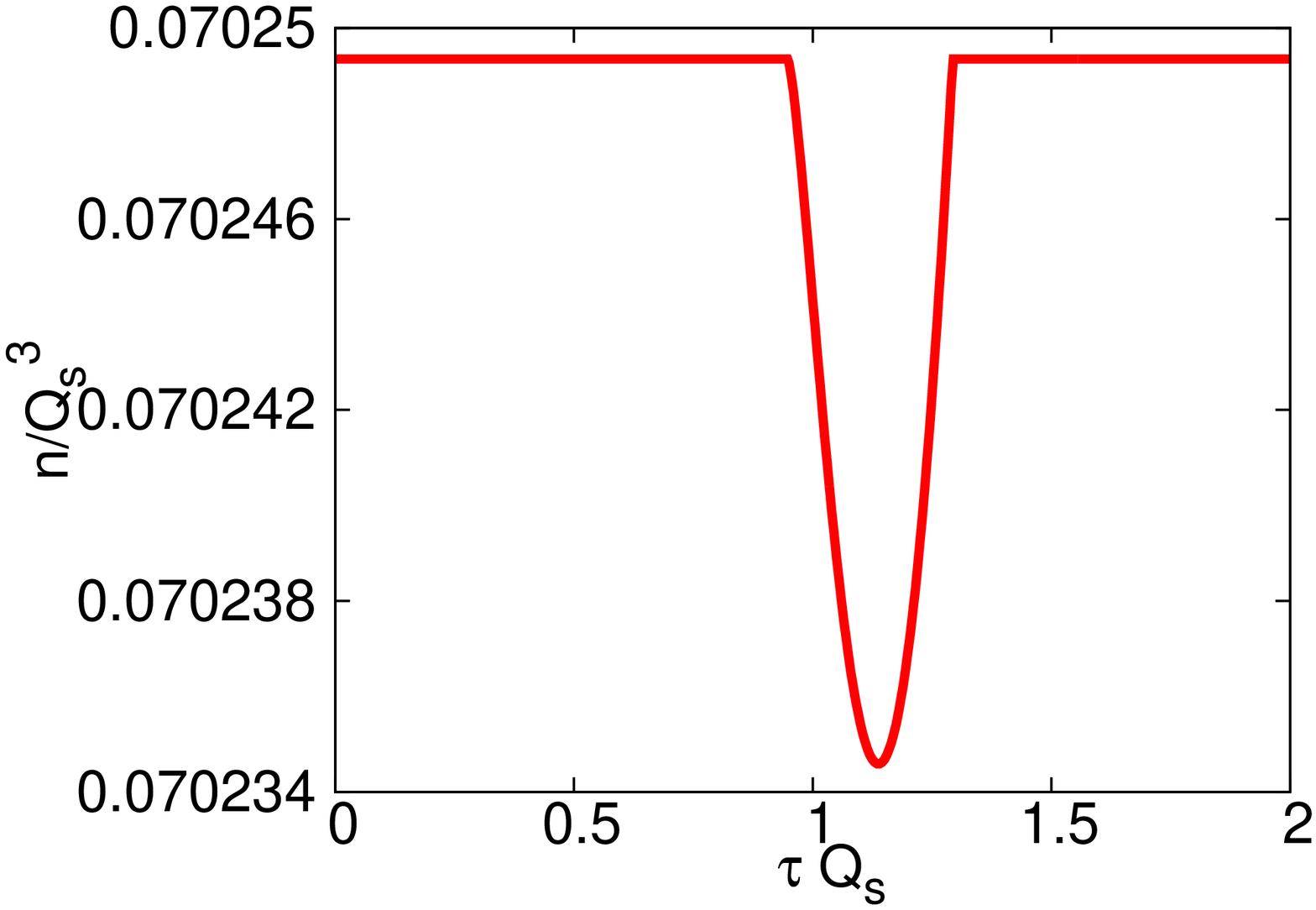}\hspace{0.05\textwidth}
\includegraphics[width=0.45\textwidth]{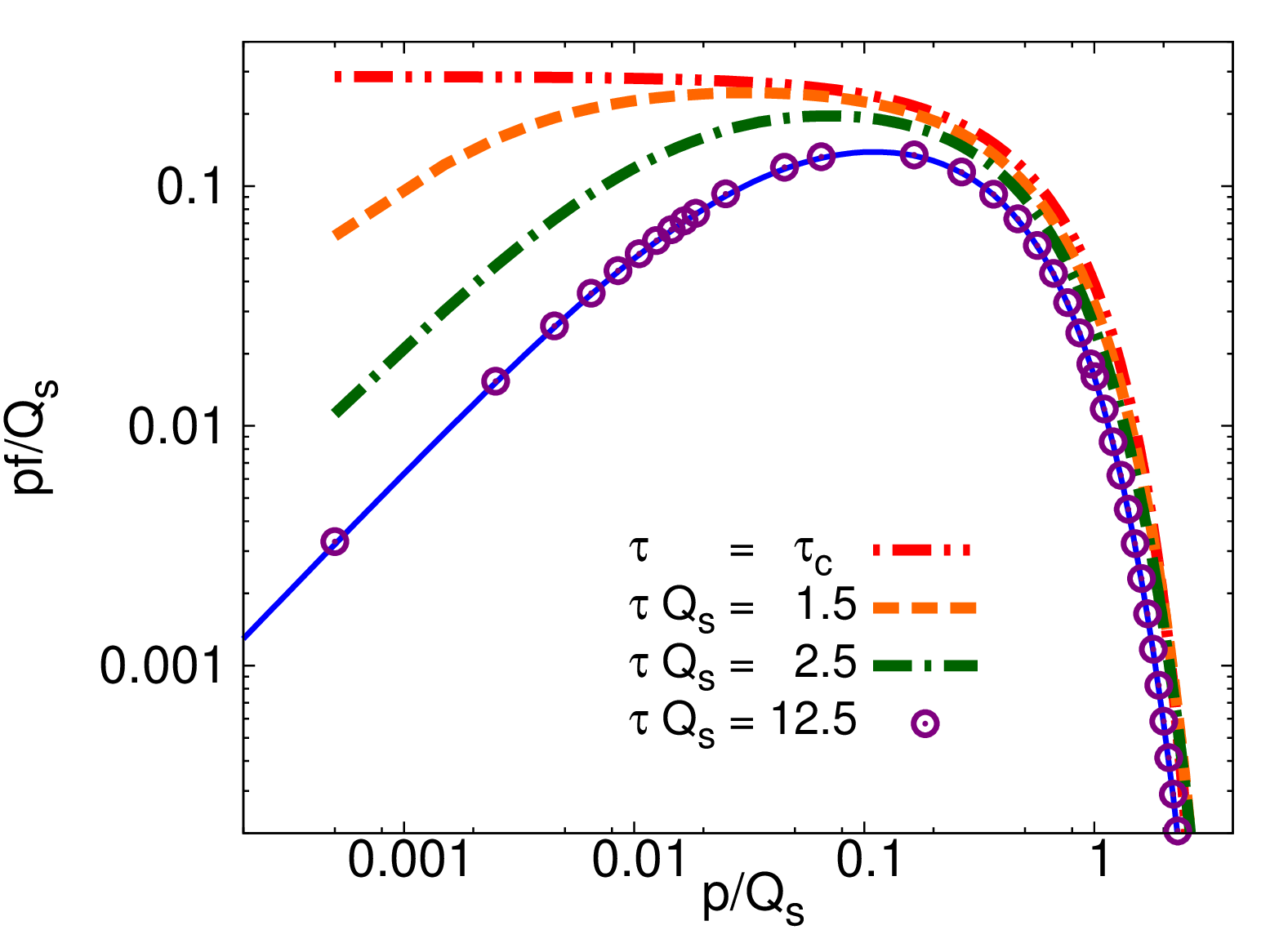}
\end{center}
\caption{(Color online)  Thermalization with transient BEC. The system with $f_0 = 0.26$ and $N_f = 3$ serves as an example of thermalization with transient BEC. Left panel shows $n$ as a function of $\tau$. BEC starts to be formed at $\tau_c = 0.947~Q_s^{-1}$ but the condensates exist only in a short period $\Delta \tau \simeq 0.35~Q_s^{-1}$. Right panel shows how the gluon distribution $f$ evolves into a thermal distribution after $\tau=\tau_c$. The solid blue curve is the thermal distribution in eq. (\ref{eq:feq}) with $T = 0.250~Q_s$ and $\mu = -0.0357~Q_s$.}\label{fig:TransientBEC}
\end{figure}

This BEC can only exist for a finite period of time since thermodynamics tell us that the system should evolve into thermal 
equilibrium without BEC.  As shown in the left panel of Fig. \ref{fig:TransientBEC}, $n$ starts to decrease at $\tau = \tau_c$, 
which indicates the formation of BEC. However, $n$ restores its original value after a period $\Delta \tau \simeq 0.35~Q_s^{-1}$. 
Afterwards, the solution with vanishing $\left.\mathcal{F}_g\right|_{p=0}$ exists again, which describes the subsequent evolution of 
the system. From that point on $n$ does not change anymore. As 
expected, $f$, as well as $F$, eventually becomes  thermal distributions, which is shown in the right panel of Fig. \ref{fig:TransientBEC}. 
For an even larger $f_0$, the transient BEC exists for a longer time. For example, when $f_0 = 0.28$, we find that BEC starts to form at $\tau_c = 0.52~Q_s^{-1}$ 
and exists for a period of $\Delta \tau \simeq 2.80~Q_s^{-1}$. In summary, the system with $f_{0c} > f_0 > f_{0t}$ serves 
as an example of thermalization with the formation of a transient BEC.

\section{Discussions}\label{sec:discussions}

In this paper we have studied the thermalization of a spatially homogeneous quark-gluon plasma, starting from an initial dense system of gluons. 
Two coupled transport equations 
for the gluon distribution $f$, and the quark distribution $F$, 
have been derived using  
the diffusion 
approximation of the Boltzmann equation, with the collision term accounting for all possible $2\leftrightarrow 2$ scatterings between quarks and gluons. 
These transport equations are solved numerically to study how the system evolves from an initial gluon distribution 
$f_0~\theta(1-\f{p}{Q_s})$ into a thermalized state of the quark-gluon plasma. We have studied systems with different values of 
$f_0$. $N_f$, the number of flavors of quarks that can be taken as massless, is also taken as a free parameter  to study the influence 
of quark production on the formation of BEC and the equilibration process (more precisely, we compare the situation where $N_f=3$ to that where $N_f=0$). 
Our main conclusions are

\begin{itemize}
\item Quark production slows down the growth of $f$ at $p\ll Q_s$.\\
For $N_f=0$, a BEC 
forms for $f_0>\left. f_{0c}\right|_{N_f = 0}=0.154$ in agreement with Ref. \cite{Blaizot:2013:BEC}. For  finite $N_f$, there is a range of values of $f_0$ larger than $\left. f_{0c}\right|_{N_f = 0}$ for which quark production hinders the formation of a BEC, and for which  the system thermalizes without the formation of a BEC. This occurs for $f_{0c}\leq f_0 \leq f_{0t}$, where 
 $f_{0t}$ depends on $N_f$. We find $f_{0t}\simeq 0.25937$ for $N_f = 3$. 
\item A transient BEC may develop in intermediate stages prior thermalization.\\
The critical value $f_{0c}$ characterizing overpopulation depends on $N_f$. $f_{0c} = 0.308$ for $N_f = 3$. A  BEC is not expected to be formed in equilibrium when $f_0 < f_{0c}$. However, we find that a transient BEC appears whenever  
 $f_{0c}>f_0 > f_{0t}$. This is a consequence of the transient excess of low momentum gluons: the growth of low momentum gluon modes is a rapid process, while quark production is relatively much slower.  The condensate only 
exists for a short period of time before quark production eventually takes over and suppresses the excess gluons as the system approaches thermal equilibration.
\item In the regime of large overpopulation, i.e. for $f_0 \gtrsim 1$,  the formation of BEC occurs at a finite  time $t_c$ given by the simple formula $t_c \sim \f{1}{(\alpha_s f_0)^2}\f{1}{Q_s}$. $t_c$ is (almost) independent of $N_f$, that is, when $f_0$ is large enough, the onset of BEC is not affected by quark production.

\item Quark production delays thermalization, and quarks are produced dominantly at small momenta.
The equilibration time, defined in eqs. (\ref{eq:teqI}) and (\ref{eq:teqII}),  is typically about 5 to 6 times larger for $N_f = 3$ than 
that for $N_f = 0$. 

\end{itemize}

The later observations may have interesting  phenomenological consequences, in particular on soft electromagnetic signals \cite{Chiu:2012ij}, or the elliptic flow \cite{Ruggieri:2013bda}. However, independently of such potential phenomenological applications, there remain  several important theoretical issues that are not addressed in this paper, and that need to be addressed. Like in Refs. \cite{Epelbaum:2011pc, Blaizot:2013:BEC}, 
we only focus on the thermalization of a spatially homogenous non-expanding system. The formation of BEC may also occur in the 
expanding quark-gluon system\cite{Blaizot:2011xf}. It would be of great interest to extend the present work to, say, the boost-invariant $1+1$ dimensional expanding system\cite{Mueller:1999:Boltzmann}. 
Moreover, the inelastic processes such as $2\leftrightarrow3$ are ignored in our transport equations, and it would be important to study how these modify the physical picture that emerges from the present calculation \cite{Huang:2013:2to3}. Besides,  
all the partons are taken as massless and like Ref.s \cite{Mueller:1999:Boltzmann,Venugopalan:2000:Thermalization,Blaizot:2013:BEC} the 
diffusion approximation is used to simplify the Boltzmann equation. The evolution of the condensates is simply described here 
 by properly added boundary conditions. It would be important to 
check how reliable those approximations are by a more elaborated investigation on how the low momentum gluons evolve over time\cite{condensate}. Finally, the validity of the kinetic description, although widely used in this type of problems, needs to be checked against the statistical classical field simulations, which may be more appropriate at early times \cite{Epelbaum:2011pc, Mueller:Son:2002}. Comparison of the present kinetic approach with the recent studies (see for instance \cite{Berges:2013:Turbulent, Gelis:2013:Isotropization} and references therein) would be particularly relevant. We leave all those interesting issues for future studies.

\section*{Acknowledgements}
We would like to thank F. Gelis for many illuminating discussions. In addition, JPB thanks J. Liao and L. McLerran for collaboration that benefited this work. BW is supported by the Agence Nationale de la Recherche project \# 11-BS04-015-01.
The research of JPB and LY is supported
by the European Research Council under the Advanced
Investigator Grant ERC-AD-267258.

\appendix

\section{Diffusion approximation of the collision integral}\label{app:M2}

\begin{table}
\begin{center}
\setlength{\tabcolsep}{.01\textheight}
      \begin{tabular}{ | l | c  | c  |}
      \hline
      $ab\leftrightarrow cd$  & $ | {\mathcal M}_{cd}^{ab}|^2/g^4 $ & In diffusion approximation
\\ \hline
%
%
$q_1 q_2\leftrightarrow q_1q_2$&
\multirow{4}{*}{$4 N_c C_F \left(\frac{s^2+u^2}{t^2}\right)$}&\multirow{4}{*}{$8 N_c C_F \frac{s^2}{t^2}$}\\
	$q_1\bar q_2\leftrightarrow q_1\bar q_2$&& \\
	$\bar q_1 q_2\leftrightarrow \bar q_1 q_2$&&\\
	$\bar q_1 \bar q_2\leftrightarrow\bar q_1\bar q_2$&&
\\\hline
%
%
	$q_1q_1\leftrightarrow q_1q_1 $ &\multirow{2}{*}{$4 N_c C_F\left(\frac{s^2+u^2}{t^2}+\frac{s^2+t^2}{u^2}\right)
      -8 C_F \frac{s^2}{tu}$}& \multirow{2}{*}{$ 8 N_c C_F \left( \frac{s^2}{t^2} + \frac{s^2}{u^2} \right)$}\\
     $\bar q_1\bar q_1\leftrightarrow\bar q_1\bar q_1$ &&
\\ \hline
%
%
      $q_1\bar q_1\leftrightarrow q_1\bar q_1$&
	$4 N_c C_F\left(\frac{s^2+u^2}{t^2}+\frac{t^2+u^2}{s^2}\right)- 8 C_F\frac{u^2}{st}$& 
	$8 N_c C_F \frac{s^2}{t^2}$\\ 
\hline
%
%
      $q_1 \bar q_1\leftrightarrow q_2\bar q_2$ & $4 N_c C_F\left(\frac{t^2+u^2}{s^2}\right)$ & 0 \\ 
\hline
%
%
      $q_1\bar q_1\leftrightarrow gg$ & $8 N_c C_F^2\left(\frac{u}{t}+\frac{t}{u}\right)- 8 N_c^2 C_F\frac{t^2+u^2}{s^2}$&
 	$- 8 N_c C_F^2 \left( \frac{s}{t} + \frac{s}{u} \right)$
\\ \hline
%
%
	$q_1 g\leftrightarrow q_1 g$ & 
	\multirow{2}{*}{$-8 N_c C_F^2\left(\frac{u}{s}+\frac{s}{u}\right)+ 8 N_c^2 C_F\frac{u^2+s^2}{t^2}$}&
	\multirow{2}{*}{$-8 N_c C_F^2 \frac{s}{u} + 16 N_c^2 C_F\frac{s^2}{t^2}$}\\
	$\bar q_1 g\leftrightarrow \bar q_1 g$ &&\\ 
	\hline
%
%
      $gg\leftrightarrow gg$&
	$ 16 N_c^2 (N_c^2 - 1)\left(3-\frac{su}{t^2}-\frac{st}{u^2}-\frac{tu}{s^2}\right)$& 
	$ 16 N_c^2 (N_c^2 - 1) \left( \frac{s^2}{t^2} + \frac{s^2}{u^2} \right)$\\\hline
      \end{tabular}
     \caption{ \label{qcd2t2}Squares of the $2\leftrightarrow2$ scattering amplitudes in QCD, with spins and colors of all four partons summed over. The dominant contributions of each process in diffusion approximation are given in the third column.  The terms proportional to $\f{s^2}{t^2}$ or $\f{s^2}{u^2}$ contribute to the diffusion currents while the terms proportional to $\f{s}{t}$ or $\f{s}{u}$ only contribute to the source terms. Here, $q_1$ ($\bar q_1$) and $q_2$  ($\bar q_2$) represent quarks (antiquarks) of different flavors.}
\end{center}
\end{table}

In this appendix we simplify the collision term of the Boltzmann equation in  \Eq{collision} within the diffusion approximation\cite{PhysicalKinetics}. 
The squares of the amplitudes for all the $2\leftrightarrow2$ processes in QCD are listed in \Tab{qcd2t2}. The momenta of the partons in the final state 
of these scattering processes are denoted respectively by $K$ and $K'$. We only need to keep all the dominant contributions in the limit that the momentum 
transfer $Q$ is much smaller than the momenta of the two scattering partons, which are denoted respectively by $P$ and $P'$. Let us take the $t$ channel 
dominated processes as an example, in which case  $Q = K - P$. In the diffusion limit, the Mandelstam variables reduce to
\begin{subequations}
\bea
s&=&(P+P')^2 = 2pp'-2{\bf p}\cdot {\bf p}'=2pp'(1-{\bf v}\cdot {\bf v}')\,,\\
t&=&Q^2 \simeq -q^2+({\bf q}\cdot {\bf v})^2\,,\\
u&=&(P-K')^2 \simeq -2pp'(1-{\bf v}\cdot {\bf v}') = -s
\ea
\end{subequations}
with ${\bf v}\equiv{\bf p}/p$ and ${\bf v'}\equiv{\bf p'}/p'$, and
\be
\delta(E_p + E_{p'} - E_k - E_{k'} )\simeq \delta(\bf{q}\cdot(\bf{v}' - \bf{v})).
\ee
The corresponding contributions from the $u$ channel scattering can be obtained by simply interchanging $K$ and $K'$.  The leading contributions to 
$|{\mathcal M}_{cd}^{ab}|^2$ in the small angle approximation are given in the third column of Table \ref{qcd2t2}. By plugging the terms proportional to $\f{s}{t}$ or $\f{s}{u}$ into 
the collision term of eq. (\ref{collision}), one can easily obtain the source terms
\bea
S_g &=& -\f{N_f}{C_F} S_q = \frac{2 \alpha_s^2 N_f  C_F}{p} \left[ {F_{\bf p}} (1+f_{\bf p})-{f_{\bf p}} (1 - {F_{\bf p}})) \right]\nn
&&\times \int \frac{d^3\p'}{(2\pi)^3} \f{1}{{p'}}({f_{\bf p'}}+{F_{\bf p'}}) \int  d^3{\bf q} \frac{1-{\bf v \cdot \bf v'}}{  q^2 - (\bf v \cdot \bf q)^2} \delta(\bf{q}\cdot(\bf{v}' - \bf{v}))\nn
&=&\frac{4\pi\alpha_s^2\L C_FN_f\I_c}{ p}\left[ F_{\bf p}(1+f_{\bf p}) - f_{\bf p} ( 1-F_{\bf p} ) \right],
\ee
where we have used the integral
\bea
\label{logint1}
&&\int d^3{\bf q} \left[\frac{1-{\bf v\cdot \bf v'}}{q^2-({\bf q\cdot \bf v})^2}\right]\delta ({\bf q\cdot(\bf v-\bf v')})
=2\pi\L.
\ee
The terms of $| \mathcal M_{cd}^{ab}|^2$ proportional to $\f{s^2}{t^2}$ and $\f{s^2}{u^2}$ in the limit $q\ll p, p'$ only contribute to the diffusion terms in the collision term of the transport equations. Let us write
\bea\label{eq:Cdiff}
\C[f_{\bf p}^a]=&&\frac{1}{2 p }\sum\limits_{b,c,d}\int \frac{d^3\p'}{(2\pi)^3}
\frac{d^3{\bf q}}{(2\pi)^3} w^{ab}_{cd}\left(\bf p +\f{\bf q}{2},\bf p'-\f{\bf q}{2}, \bf q \right) \nonumber\\
&&\times\left[f_{|\bf p + \bf q|}^c f_{ |\bf p' - \bf q| }^d(1 + \epsilon_a f_{\bf p}^a)(1 + \epsilon_b f_{\bf p'}^b) - f_{\bf p}^a f_{\bf p'}^b (1 + \epsilon_c f_{|\bf p + \bf q|}^c)(1 + \epsilon_d f_{ |\bf p' - \bf q| }^d))\right],
\ea
where the relation of $w^{ab}_{cd}$ to $|{\mathcal M}_{cd}^{ab}|^2$ can be obtained by referring to eq. (\ref{collision}),  $\epsilon_i = 1$ for gluons and $\epsilon_i = -1$ for quarks and antiquark.
To derive the diffusion terms of the transport equations, one only needs to keep the terms in which the factors in the parentheses $[\cdots]$ on the right hand side of 
eq. (\ref{eq:Cdiff}) vanish in the limit $\bf q \to 0$.  In this case, partons $c$ and $d$ can be respectively taken as the same species as $a$ and $b$
\footnote{Here, we need only to consider the dominant terms from the $t$ channels. There are equal contributions from the $u$ channels if particles $c$ and $d$ are identical 
particles. However, the sum of the contributions from both channels should be divided by  $2$.}. Therefore, the diffusion terms describe the diffusion of particle $a$ in 
the momentum space as a result of scattering off particle $b$. They are different from the source terms, which are proportional to the production rate of particle $b$ of a 
different species from the scattering parton $a$ with another parton.  By expanding the integrand of eq. (\ref{eq:Cdiff}) in powers of $q$ and keeping only the first non-vanishing 
term, we find, after some algebra,
\bea\label{eq:Cdiffofq}
\C_{diff}[f_{\bf p}^a]&&=-\nabla_{\bf p} \J^a,
\ea
where the diffusion current for particle $a$ is given by
\bea\label{eq:Jdef}
 {\J^a}^i&\equiv& - \frac{1}{2 p } \sum\limits_{b}\int \frac{d^3\p'}{(2\pi)^3} \frac{d^3{\bf q}}{(2\pi)^3}\, w_{diff}^{ab}\left(\bf p ,\bf p', \bf q \right) \nn
 &&\times\f{q^i q^j}{2} \left[ f^b_{\bf p'} (1+ \epsilon_b f^b_{\bf p'}) \nabla_{p^j} f_{\bf p}^a - f_{\bf p}^a (1+ \epsilon_a f_{\bf p}^a) \nabla_{p'^j} f^b_{\bf p'} \right]
\ee
with
\bea
w_{diff}^{ab}\left(\bf p ,\bf p', \bf q \right) = \f{1}{8 p {p'}^2 \nu_a} 2 \pi \delta(\vec{q}\cdot({\bf v}' - {\bf v})) | {\mathcal M}_{ab}^{ab}|^2_{diff}.
\ee
Here, $| {\mathcal M}_{ab}^{ab}|^2_{diff}$ are the terms proportional to $\f{s^2}{t^2}$ in the third column of Table \ref{qcd2t2}. To simplify $\J^a$, we need to evaluate
\bea
\label{logint2}
B^{ij} &&\equiv \int \frac{d^3{\bf q}}{(2\pi)^3} \frac{q^i q^j(1-{\bf v\cdot \bf v'})^2}{[q^2-({\bf q\cdot \bf v})^2]^2} 2 \pi \delta ({\bf q\cdot(\bf v-\bf v')})\nn\\
&& =\f{\L}{4\pi}[\delta^{ij}(1-\v\cdot\v')+(v^iv'^{j}-v'^{i}v^j)],
\ee
and
\bea
J^{ab} &&\equiv - \f{g^4}{8\nu_a} \nabla_{p^i} \int \frac{d^3\p'}{(2\pi)^3} B^{ij}\left[ f^b_{\bf p'} (1+ \epsilon_b f^b_{\bf p'}) \nabla_{p^j} f_{\bf p}^a - f_{\bf p}^a (1+ \epsilon_a f_{\bf p}^a) \nabla_{p'^j} f^b_{\bf p'} \right]\nn
&&= - \f{ \pi \alpha_s^2 \L}{2 \nu_a} \nabla_{\bf p} \cdot \int \frac{d^3\p'}{(2\pi)^3} \left[ f^b_{\bf p'} (1+ \epsilon_b f^b_{\bf p'}) \nabla_{\bf p} f_{\bf p}^a + \f{2 f^b_{\bf p'}}{p'} f_{\bf p}^a (1+ \epsilon_a f_{\bf p}^a) \bf v \right].
\ee
Here, we have assumed that $f^b_{\bf p} = f^b_{- \bf p}$ in order to get $J^{ab}$ in the last line in the above equation. $\J^a$ in eq. (\ref{qg_currents}) is obtained by summing $J^{ab}$ over $b$ with the coefficient given by that of the corresponding term proportional to $\f{s^2}{t^2}$ in the third column of Table \ref{qcd2t2}.
\section{Series solutions and boundary conditions to the transport equations}\label{app:series}
%
%
As discussed in the main text, there are two types of solutions of the transport equations, characterized by
the behavior of the gluon distribution near the origin $p=0$: either $f(p=0)$ is a finite constant, or $f(p\rightarrow 0)\sim 1/p$.
In order to analyze further these solutions, we set
\be
f=\sum\limits_{n} c_{n} p^n, \qquad F=\sum\limits_{n} d_{n} p^n
\ee
where the coefficients $c_{n}$ and $d_{n}$ can be determined from the transport equations in eqs. (\ref{eq:ftoSolve}) 
and (\ref{eq:FtoSolve}) with $I_a$, $I_b$ and $I_c$  functions of $\tau$. One then finds that there are only 
two types of solutions allowed by the transport equations:
\begin{itemize}
\item $f$ is analytic at $p = 0$.\\ 
In this case, we have
\bea
f&=&c_{0}+\frac{C_{F} N_{f} I_{c} [c_{0}-(2 c_{0}+1) d_{0}]-2 N_{c} c_{0} (c_{0}+1) I_{b}}{2 N_{c} I_{a}} p+O(p^2),\nn
F&=&d_{0}+\frac{C_{F} I_{c} [c_{0} (2 d_{0}-1)+d_{0}]+2 (d_{0}-1) d_{0} I_{b}}{2 I_{a}} p+ O(p^2),
\ea
and
\bea\label{currentssmallp}
-J_g&=&-\frac{C_{F} N_{f} I_{c} [c_{0} (2 d_{0}-1)+d_{0}]}{2 N_{c}} + O(p)
,\nn
-J_q&=&\frac{1}{2} C_{F} I_{c} [ c_{0} (2 d_{0}-1)+d_{0}]+O(p)
\ee
with $c_0 = f(\tau, 0)$ and $ d_0 = F(\tau, 0)$.

In the limit $c_0 \gg 1$, the radius of convergence of the above series solution shrinks to zero. In this case, we find
\bea
&&f= c_0 + \left[ - c_0^2 + O(c_0) \right]\f{p}{T^*}  + \left[ c_0^3 + O(c_0^2) \right]\left( \f{p}{T^*} \right)^2 +\cdots,\
\ee
with $T^* \equiv \f{I_a}{I_b}$. The leading terms in $c_0$ at each order in $p$ can be resummed and, thus, we obtain
\be
f\simeq \f{c_0}{1+ \f{c_0 p}{T^*}} = \f{T^*}{p - \mu^*_g}.
\ee
This is the classical distribution function, with an effective chemical potential given $\mu^*_g \equiv - T^*/c_0$. 
The above resummed solution is very useful for understanding the evolution of the quark-gluon system close to $\tau_c$\cite{Blaizot:2013:BEC}.
\item $f$ is singular at $p = 0$.\\ 
In this case, we get
\bea
f&=&\frac{c_{-1}}{p}-\frac{1}{2}\nn
&&+\frac{ I_{a} \left[-C_{F} N_{f} I_{c}+2 N_{c} \dot c_{-1}+N_{c} I_{b}\right]+C_{F} c_{-1} I_{c} \left[I_{b} (N_{f}-N_{c})-2 N_{c} \dot c_{-1}\right]}{4 N_{c} (2 c_{-1} I_{b}+I_{a}) (I_{a}-C_{F} c_{-1} I_{c})} p + O(p^2),\nn
F&=&\frac{1}{2}+\frac{ I_{b}-C_{F} I_{c}}{4 ( C_{F} c_{-1} I_{c} - I_{a} )} p+O(p^3),
\ea
and
\bea
-J_g&=&\frac{c_{-1} (c_{-1} I_{b}-I_{a})}{p^2}+\frac{1}{4} \left(\frac{C_{F} N_{f} I_{c} (I_{a}-c_{-1} I_{b})}{N_{c} (C_{F} c_{-1} I_{c}-I_{a})}+2 \dot c_{-1}\right)+ O(p^2),\nn
-J_q&=&\frac{C_{F} I_{c} (I_{a}-c_{-1} I_{b})}{4 (I_{a}-C_{F} c_{-1} I_{c})}+O(p^2).
\ea
\end{itemize}

To solve the transport equations in eqs. (\ref{eq:ftoSolve}) and (\ref{eq:FtoSolve}), one needs two initial conditions and four boundary conditions. In our code, we use the following boundary conditions
\bea\label{eq:bcnumerics}
&&\left.\mathcal{F}_g\right|_{p=\infty} = 0,~~ \left.\mathcal{F}_g\right|_{p=0}= 4\pi c_{-1}\left( I_a - I_{b}c_{-1} \right),~~ \left.\mathcal{F}_q\right|_{p=\infty} = 0,~~ \left.\mathcal{F}_q\right|_{p=0}=0.
\ea
The explicit Euler method is used for time integration and only $c_{-1}$ at the current time step is needed for the calculation of $f$ and $F$ at the next time step.


\end{document}